\documentclass[10pt,aps,prl,twocolumn,preprintnumbers,amsmath,amssymb,floatfix,showpacs,citeautoscript]{revtex4-1}
\usepackage[latin1]{inputenc}
\usepackage[T1]{fontenc}
\usepackage[english]{babel}
\usepackage[pdftex]{graphicx}
\usepackage{amsmath}
\usepackage{times}
\usepackage[usenames,dvipsnames,svgnames,table]{xcolor}
\usepackage[colorlinks,plainpages=false,linkcolor=blue,urlcolor=blue,citecolor=blue,pdfpagemode=UseNone]{hyperref}

\renewcommand{\AA}{\text{\r{A}}}

\begin{document}

\title
{
\boldmath
Active learning and element embedding approach in neural networks\\for infinite-layer versus perovskite oxides
}

\author{Armin Sahinovic}
\affiliation{Department of Physics, Universit\"at Duisburg-Essen, Lotharstr.~1, 47057 Duisburg, Germany}
\author{Benjamin Geisler}
\email{benjamin.geisler@uni-due.de}
\affiliation{Department of Physics, Universit\"at Duisburg-Essen, Lotharstr.~1, 47057 Duisburg, Germany}

\date{\today}

\begin{abstract}
Combining density functional theory simulations and active learning of neural networks, we explore formation energies of oxygen vacancy layers, lattice parameters, and their correlations in infinite-layer versus perovskite oxides across the periodic table,
and place the superconducting nickelate and cuprate families in a comprehensive statistical context.
We show that neural networks predict these observables with high precision,
using only $30$-$50\%$ of the data for training.
Element embedding 
autonomously identifies concepts of chemical similarity between the individual elements in line with human knowledge.
Based on the fundamental concepts of entropy and information,
active learning composes the training set by an optimal strategy without \textit{a priori} knowledge
and provides systematic control over the prediction accuracy.
This offers key ingredients to considerably accelerate scans of large parameter spaces
and exemplifies how artificial intelligence
may assist on the quantum scale in finding novel materials with optimized properties. \end{abstract}


\maketitle


Over the last years, artificial intelligence (AI) algorithms have attracted increasing attention in computational materials science.
Machine learning techniques~\cite{ButlerWalsh:18, SchmidtMarques:19, WangSparks:20, WardWolverton:16, FaberLilienfeldArmiento-Elpasolite:16, Ghiringhelli-Descriptors:15, Bartel-ToleranceFactor:19, SchmidtMarques:17, Viewpoint:19, LopezLittlewood:20}
such as deep learning~\cite{XieGrossman:18, YeOng:18, JhaAgrawal:19, AgrawalChoudhary:19}
allow for a variety of different intriguing and often unconventional approaches, 
ranging from applications in molecular dynamics~\cite{DL-MD-Car:18},
the unsupervised identification of latent knowledge in scientific literature~\cite{TshitoyanJain-UnsupervisedWordEmbedding:19},
to the understanding of chemical trends from materials data~\cite{Atom2VecPNAS:18, JhaWolvertonAgrawal:18}.
%
%
In parallel,
the increasing computational resources have driven high-throughput searches
to identify novel materials with enhanced properties,
which resulted in the emergence of different materials databases~\cite{MatProj:13, OQMD:13, AFLOW:12, NOMAD:18}.
However, screening large parameter spaces by quantum-scale materials simulations,
e.g., employing density functional theory (DFT),
is still impeded by a high energy and time consumption.

Aiming for a more efficient strategy,
here we complement systematic first-principles simulations across the periodic table with deep learning of artificial neural networks (NNs).
We use the topical infinite-layer oxides (IL, $AB$O$_2$)~\cite{Li-Supercond-Inf-NNO-STO:19, Nomura-Inf-NNO:19, JiangZhong-InfNickelates:19, Sakakibara:19, JiangBerciuSawatzky:19, Botana-Inf-Nickelates:19, Osada-PrNiO2-SC:20, Lechermann-Inf:20, NNO-SC-Thomale:20, NNO-SelfDopingDesign-d9-Arita:20, Kitatani-AritaZhongHeld:20, GeislerPentcheva-InfNNO:20, Li-Supercond-Dome-Inf-NNO-STO:20, Choi-Lee-Pickett-4fNNO:20, Si-Zhonh-Held:InfNNO-Hydrogen:20, GeislerPentcheva-NNOCCOSTO:21, Ortiz-NNO:21}
and the respective perovskites (P, $AB$O$_3$)
to show that NNs are capable of understanding
the formation energies of oxygen vacancy layers,
as well as the lattice parameters of the individual compounds.
These observables act as a fingerprint of the reduction reaction.
Hence, despite the complexity of these two materials classes and their relations,
as evidenced by detailed statistical analysis,
NNs autonomously unravel the systematics of their quantum-chemical bonding 
by using just $30$-$50\%$ of the data for training.
Subsequently, they predict the properties of all compounds, even those they have never seen, with high accuracy, well within the error bars of DFT itself.
%
Interestingly, it turns out to be sufficient to only provide the $A$- and $B$-site element names as input to the NNs, and no further atomic properties.
Element embedding~\cite{Atom2VecPNAS:18,Word2Vec:13} leads to the emergence of a very unique AI understanding of the chemical relations between the individual elements that mirrors the conventional picture of the periodic table.
Finally, we show that combining these techniques with active learning~\cite{SchmidtMarques:19, Lookman-AL:19}
allows for an efficient screening of the materials parameter space,
being clearly superior to a randomly selected training set and providing systematic accuracy control.
We provide detailed visual insight into the algorithm's working mechanisms and its performance,
exemplifying the potential of AI to considerably accelerate high-throughput materials optimization.

\textit{Methodology.}
We performed first-principles simulations~\cite{KoSh65, USPP-PAW:99, PAW:94, PeBu96}
to construct a database of ground-state energies and optimized lattice parameters for $4692$ combinations of different elements at the $A$ and $B$ sites (as detailed below) for both the P and the IL oxides, 
which were modeled by using cubic~\cite{SchmidtMarques:17} and tetragonal~\cite{Botana-Inf-Nickelates:19, Lechermann-Inf:20} unit cells, respectively.
We adopted the DFT$+U$ standards of the Materials Project database~\cite{MatProj:13, PYMATGEN:13, LiechtensteinAnisimov:95}.
As a difference, rare-earth $4f$ electrons were consistently frozen in the core~\cite{Liu-NNO:13, Nomura-Inf-NNO:19, Lechermann-Inf:20, GeislerPentcheva-InfNNO:20, GeislerPentcheva-NNOCCOSTO:21}.
For the elemental bulk references, we used the Materials Project ground state crystal structures and energies, recalculating $4f$ and finite-$U$ compounds to ensure consistency.
NNs were realized in Keras/Tensorflow~2~\cite{Tensorflow:15, CholletKeras:15},
and the active-learning algorithm was developed in Python~3.
The formation energies of the oxygen vacancy layers are determined from DFT ground-state energies by
$E_f^{V_\text{O}} = E(AB\text{O}_2) - E(AB\text{O}_3) + \mu_\text{O}^{}$,
where $\mu_\text{O} = \frac{1}{2} E(\text{O$_2$})$ models the oxygen-rich limit
\footnote{The well-known overbinding of gas-phase O$_2$ molecules in DFT necessitates a correction of $E(\text{O$_2$})$,
which we performed such as to reproduce the experimental O$_2$ binding energy of $5.16$~eV~\cite{GeislerPentcheva-LNOLAO-Resonances:19, GeislerPentcheva-LCO:20, LNO-OxVac-Beigi:15}.}.
The heats of formation of the P phase from the constituent bulk elements read
$E_f^\text{P} = E(AB\text{O}_3) - E(A~\text{bulk}) - E(B~\text{bulk}) - 3 \mu_\text{O}^{}$.
All energies are given per formula unit.



\begin{figure}
	\centering
	\includegraphics[]{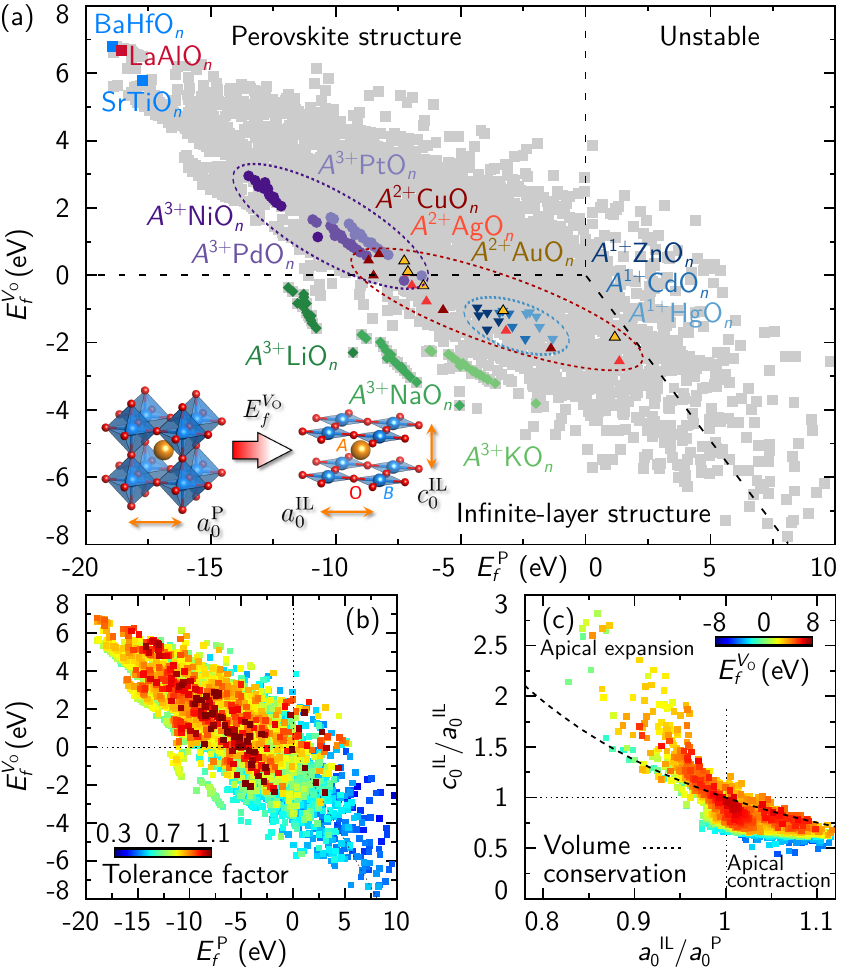}
	\caption{(a)~Reduction of the apical oxygen ions in P oxides ($n=3$) results in the emergence of the anisotropic IL structure ($n=2$). This reaction is associated with the energy $E_f^{V_\text{O}}$. The phase diagram compares the relative stability of the IL versus the P structure as a function of the respective P heat of formation for the entire data set. A number of interesting compounds is highlighted. (b)~Superposition of (a) with the Goldschmidt tolerance factor~$t$. (c)~Structural perspective on the data, comparing apical to basal changes upon reduction and superimposing them with $E_f^{V_\text{O}}$.\vspace{-1.5ex}}
	\label{fig:Structures}
\end{figure}

\textit{Data exploration and statistical analysis.}
We begin by providing an overview of the data set from a thermodynamic, a structural, and a statistical perspective.
Fig.~\ref{fig:Structures}(a) displays the entire data in a $E_f^{V_\text{O}}$ vs.\ $E_f^\text{P}$ phase diagram,
comparing the relative stability of the IL and the P structure,
as well as their stability with respect to the constituent bulk elements.
This is motivated by recent experiments on IL oxides that attracted considerable attention, specifically superconducting nickelates~\cite{Li-Supercond-Inf-NNO-STO:19, Osada-PrNiO2-SC:20, Li-Supercond-Dome-Inf-NNO-STO:20},
which are initially stabilized as P films on SrTiO$_3$(001) via heteroepitaxy,
followed by a topotactic reduction of the apical oxygen ions.
$E_f^{V_\text{O}}$ ranges from $-8$ to $+7$~eV, while $E_f^\text{P}$ covers almost $30$~eV.
The plot reveals an overall linear trend, correlating the P stability and its reduction energy.
However, the data scatters broadly around the regression line $E_f^{V_\text{O}} = -0.36 \, E_f^\text{P} -1.37$~eV.
Superimposing this plot with the Goldschmidt tolerance factor $t= \frac{r_A + r_\text{O}}{\sqrt{2}(r_B+r_\text{O})}$ calculated from the ionic radii [Fig.~\ref{fig:Structures}(b)]
reflects that the P stability (moving from right to left) increases with $t$,
reducing again for $t>1$.
Again, we find that the data scatters broadly around this well-known trend.
Structural analysis [Fig.~\ref{fig:Structures}(c)] shows that
most materials exhibit the tendency to contract vertically upon reduction (up to $50\%$), expanding simultaneously in the plane (up to $10\%$) with reduced volume, particularly those materials where the reaction is exothermic ($E_f^{V_\text{O}} < 0$). 
For some very stable compounds, the changes are rather modest (center of the plot).
In sharp contrast, a few materials expand massively in apical direction ($c_0^\text{IL}/a_0^\text{IL} \sim 2-3$) with $10-20\%$ basal contraction. 

\begin{table}
	\centering
	\vspace{-1.5ex}
	\caption{\label{tab:StructEner}Energies and lattice parameters for selected systems (cf.~Fig.~\ref{fig:Structures}). $E_f^{V_\text{O}} < 0$ indicates a preferred IL structure.}
	\begin{ruledtabular}
	\begin{tabular}{lccccc}
		System & $E_f^{V_\text{O}}$ (eV) & $E_f^\text{P}$ (eV) & $a_0^\text{P}$ (\AA) & $a_0^\text{IL}$ (\AA) & $c_0^\text{IL}$ (\AA) \\
		\hline
		LaNiO$_n$ & $2.8$ & $-13.2$ & $3.82$ & $3.93$ & $3.40$ \\
		PrNiO$_n$ & $2.7$ & $-12.8$ & $3.80$ & $3.91$ & $3.35$ \\
		NdNiO$_n$ & $2.7$ & $-12.8$ & $3.79$ & $3.90$ & $3.30$ \\
		LuNiO$_n$ & $2.1$ & $-12.2$ & $3.70$ & $3.81$ & $3.02$ \\
		SrNiO$_n$ & $0.5$ & $-9.9$  & $3.83$ & $3.86$ & $3.53$ \\
		CaCuO$_n$ & $-0.01$ & $-8.5$ & $3.80$ & $3.87$ & $3.21$ \\
		SrCuO$_n$ & $0.44$ & $-8.7$ & $3.89$ & $3.95$ & $3.49$ \\
		LiZnO$_n$ & $-1.6$ & $-3.9$ & $3.86$ & $3.79$ & $3.08$ \\
		NaZnO$_n$ & $-1.2$ & $-4.3$ & $3.91$ & $3.85$ & $3.41$ \\
		LaLiO$_n$ & $-0.5$ & $-11.7$ & $3.80$ & $4.01$ & $3.26$ \\
		LaNaO$_n$ & $-1.9$ & $-8.8$ & $4.06$ & $4.27$ & $3.23$ \\
		\hline
		LaAlO$_n$ & $6.7$ & $-18.6$ & $3.81$ & $3.83$ & $3.67$ \\
		SrTiO$_n$ & $5.8$ & $-17.7$ & $3.94$ & $3.99$ & $3.59$ \\
		BaHfO$_n$ & $6.8$ & $-18.9$ & $4.20$ & $4.21$ & $4.05$ \\
	\end{tabular}
	\end{ruledtabular}
\end{table}

\begin{figure*}
	\centering
	\includegraphics[width=\textwidth]{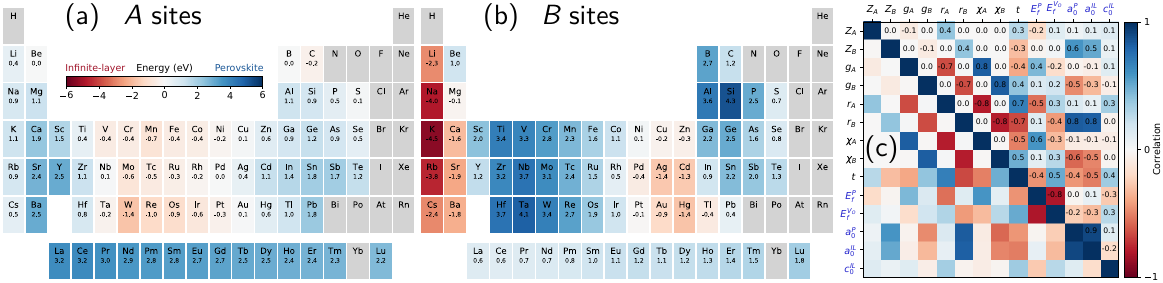}
	\caption{Statistical analysis of IL and P oxides. Panels (a) and (b) display trends of $\langle E_f^{V_\text{O}} \rangle$ across the periodic table, fixing either the $A$ or the $B$ site and subsequently averaging over the other site, indicating site-resolved which elements tend to stabilize which of the two phases. (c)~The correlation matrix unravels the interdependence of the different observables, including atomic properties (atomic number $Z$, periodic table group $g$, atomic radius $r$, and electronegativity $\chi$), the Goldschmidt tolerance factor $t$, and different energies and lattice parameters as determined from first principles (blue labels).}
	\label{fig:PSEs}
\end{figure*}

Figure~\ref{fig:Structures}(a) places the formally $d^9$ IL nickelates and cuprates in an interesting context (cf.~Table~\ref{tab:StructEner}).
%
The nickelates appear as a compact family in the phase diagram, exhibiting a stable P phase, but being simultaneously close to the IL regime;
palladates~\cite{Kitatani-AritaZhongHeld:20} and platinates are even more easily reduced.
In contrast, the cuprate family extends widely over the IL region.
This reflects the naturally preferred 4-fold coordinated plaquette structure typical for high-$T_C$ cuprate superconductors.
Continuing this series,
$d^9$ alkali-metal Zn/Cd/Hg oxides emerge, being again more compact and located deeper within the IL regime.
Further interesting compounds can be identified
in the IL region
that simultaneously exhibit a highly negative $E_f^\text{P}$. 
Exemplarily, LaLiO$_n$ and LaNaO$_n$ emerge as strongly anisotropic IL structures [Fig.~\ref{fig:Structures}(a), Table~\ref{tab:StructEner}].
They are insulators due to an $A_{}^{3+}B_{}^{1+}\text{O}_2^{2-}$ configuration
and thus may serve as quantum confinement layers.

Figures~\ref{fig:PSEs}(a) and~(b)
show averaged $\langle E_f^{V_\text{O}} \rangle$ for either a fixed $A$ or $B$ site, respectively,
unraveling site- and element-resolved trends in the relative stability of P and IL phases across the periodic table.
At the $A$ site,
most of the central transition metals induce strong tendencies towards the planar IL configuration, particularly W.
The remaining elements generally stabilize P, specifically Ca, Sr, Ba, Sc, Y, Pb, and the rare-earth metals.
We observe a decreasing trend of $\langle E_f^{V_\text{O}} \rangle$
across the rare-earth metals from $3.2$ (La) to $2.2$~eV (Lu),
and shifting from the Sc group (including rare-earth metals, $A_{}^{3+}$) to the alkali metals ($A_{}^{1+}$).
The $B$ site exhibits a much higher contrast among the different elements:
Alkali metals, particularly K, 
induce the IL phase.
The late transition metals (Ni, Cu, Zn groups) largely display an increasingly negative $\langle E_f^{V_\text{O}} \rangle$ as well,
which highlights their tendency towards the planar IL geometry discussed above [Fig.~\ref{fig:Structures}(a)].
In contrast, the P phase is clearly preferred by the early transition metals
as well as by the aluminates.
Also Si favors the formation of P oxides such as Mg$_{}^{2+}$Si$_{}^{4+}$O$_3^{2-}$ and Ca$_{}^{2+}$Si$_{}^{4+}$O$_3^{2-}$, which are abundant in the lower part of the Earth's mantle~\cite{Nestola-CaSiO3Perov:18}.

The symmetric matrix in Fig.~\ref{fig:PSEs}(c) displays the Pearson product-moment correlation coefficients between different observables,
ranging from atomic properties of the $A$ and $B$ site elements
to the energies and lattice parameters as determined from first principles.
%
$E_f^\text{P}$ shows a modest dependence on the $A$ site,
whereas $E_f^{V_\text{O}}$ lacks significant correlations apart from being anticorrelated with $E_f^\text{P}$ ($-0.8$),
which reflects the linear trend observed in Fig.~\ref{fig:Structures}.
$a_0^\text{P}$ and $a_0^\text{IL}$ correlate predominantly with the $B$ site, particularly $r_B$ ($0.8$),
and are also significantly intercorrelated ($0.9$).
In sharp contrast, $c_0^\text{IL}$ exhibits almost no correlations with the other quantities,
at most with the Goldschmidt tolerance factor~$t$.
While optimized descriptors~\cite{Bartel-ToleranceFactor:19, Ghiringhelli-Descriptors:15, Viewpoint:19} may enhance the correlation,
this indicates that a nonlinear methodology is required to reliably predict this quantity,
which turns out to be challenging, as shown below.

\begin{figure}[b]
	\centering
	\includegraphics[]{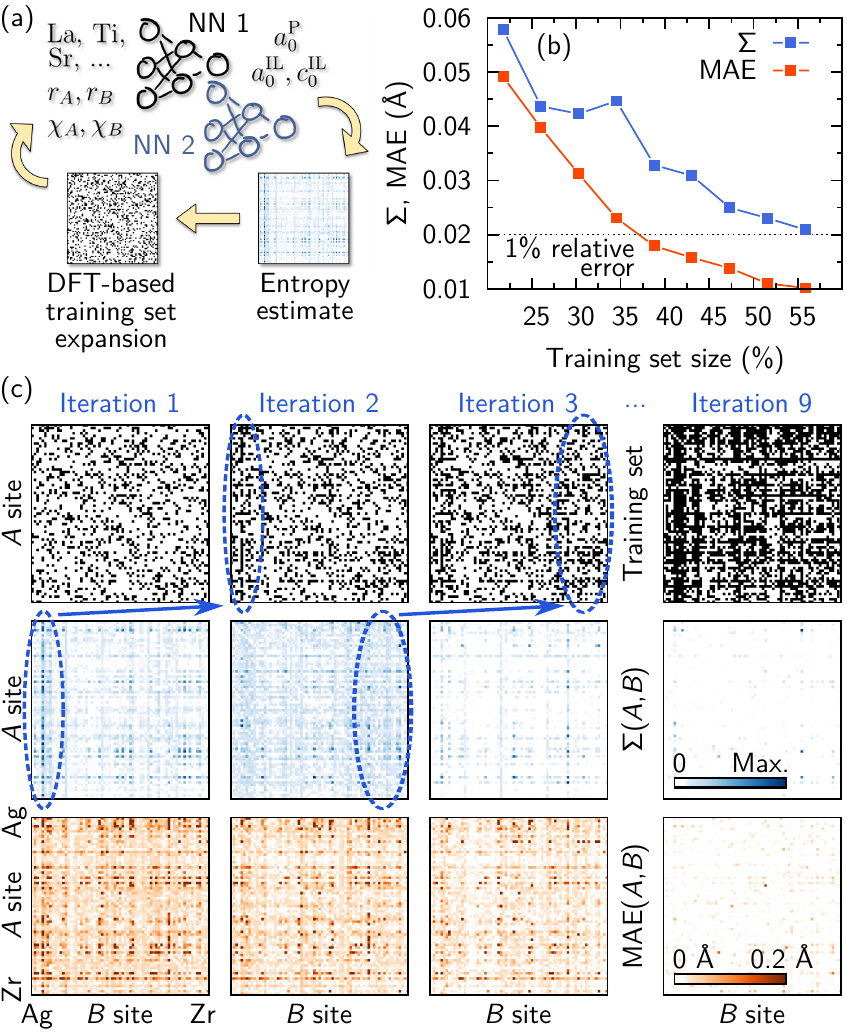}
	\caption{(a)~Active learning (AL) cycle. (b)~Evolution of $\Sigma$ and the MAE with the number of AL iterations, shown exemplarily for lattice constant prediction. (c)~Parameter space maps monitor consecutive AL iterations, the chemical elements being ordered alphabetically. For $\Sigma$ (blue), the color scale adaptively maximizes the contrast in each case, highlighting the materials selected for updating the training set (black). The MAE (red) is not used during AL, but can be exploited to trace the performance.\vspace{-1.5ex}}
	\label{fig:AL}
\end{figure}


\textit{Active learning of neural networks.}
The interesting question arises whether the insights presented so far would have been possible without explicitly calculating the entire data set, but only a fraction of it.
We address this aspect by implementing an active learning (AL) algorithm, which constitutes a form of semi-supervised learning~\cite{SchmidtMarques:19, Lookman-AL:19}.
Two NNs are trained in parallel [Fig.~\ref{fig:AL}(a)].
They take the names of the elements at the $A$ and $B$ sites as categorical input,
which are one-hot encoded and subsequently processed by a 16-dimensional embedding layer~\cite{Atom2VecPNAS:18}.
Such element embedding is inspired by word embedding~\cite{Word2Vec:13},
a technique used in language processing to represent words in a semantically insightful way in a vector space of compact dimension.
Optionally, the NNs feature a parallel numerical input channel to complement the output of the embedding layer by the atomic radii $r_{A,B}$ and the electronegativities $\chi_{A,B}$, which turned out to be largely redundant in view of the more powerful embedding technique.
This input layer is followed by a sequence of hidden layers, featuring 512, 256, and 128 densely connected neurons, respectively.
We explored different NN architectures and found the present one to yield optimal results.
The output layer provides energies or lattice parameters.
We apply error backpropagation on the training set (a small subset of the parameter space, $\sim 20$\%) to automatically adapt the weights that connect the individual neurons, until an optimal mapping from input to output is achieved.
Given the observables $x_i^{1,2}$ as predicted by NN~1 and NN~2 and the respective DFT ground truth $x_i^\text{DFT}$ (either energies or lattice parameters), we define by averaging over $i$:
\begin{eqnarray*}
\text{MAE}(A,B) & = & \left\langle \left\vert \frac{x_i^1(A,B) + x_i^2(A,B)}{2} - x_i^\text{DFT}(A,B) \right\vert \right\rangle_i \\
\Sigma(A,B) & = & \left\langle \left\vert x_i^1(A,B) - x_i^2(A,B) \right\vert \right\rangle_i
\end{eqnarray*}
Site averaging yields the mean absolute error $\text{MAE} = \langle \text{MAE}(A,B) \rangle_{A,B}$ and $\Sigma = \langle \Sigma(A,B) \rangle_{A,B}$.

In the AL cycle [Fig.~\ref{fig:AL}(a)], the training set is now updated iteratively, appending 200 materials per step that exhibit the highest $\Sigma(A,B)$, followed by further NN training.
Interestingly, this quantity represents an estimate of the local entropy in the parameter space,
which would read $H(A,B) \sim \sum_i \log \sigma_i(A,B)$
in case the predictions $x_i^{}$ followed uncorrelated normal distributions with
$\sigma_i(A,B) \sim \left\vert x_i^1(A,B) - x_i^2(A,B) \right\vert$.
%
In this spirit, the present AL algorithm statistically maximizes the information entailed in the training set.
From the definition of $\Sigma(A,B)$ it follows that the DFT ground truth is \textit{not} required by the AL algorithm
to select interesting materials candidates;
we use it only \textit{a posteriori} to analyze the AL performance [Figs.~\ref{fig:AL}(b) and (c)].

\begin{figure}
	\centering
	\includegraphics[width=8.6cm]{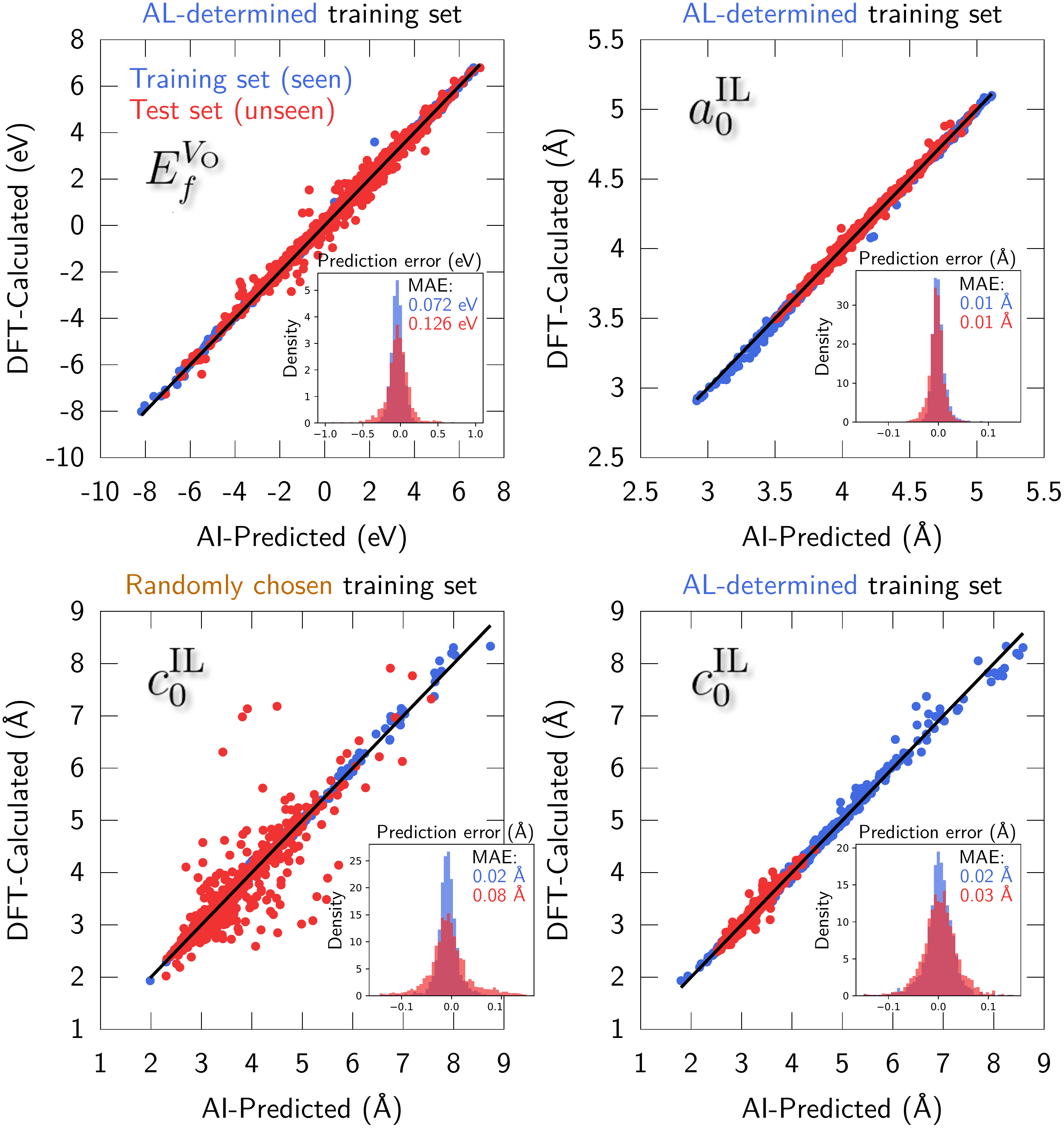}
	\caption{Prediction of $E_f^{V_\text{O}}$, $a_0^\text{IL}$, and $c_0^\text{IL}$ by a single NN versus the DFT ground truth, using only $\sim 50\%$ of the data as training set (blue points, seen by the NN). The red points represent the test set, which has never been presented to the NN before. Contrasting the AL results with those obtained for a randomly chosen training set of equal size reveals the advantages of AL. The predictions for $a_0^\text{P}$ and heats of formation are even more accurate (not shown).\vspace{-1.5ex}}
	\label{fig:AIvsDFT}
\end{figure}

\begin{figure}[b]
	\centering
	\includegraphics[width=7.0cm]{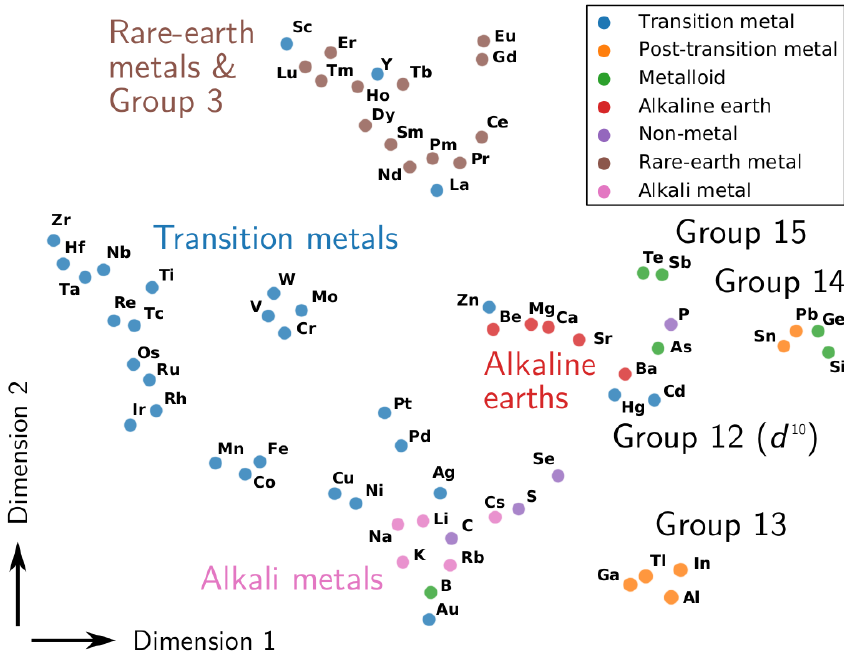}
	\caption{Stochastic neighbor embedding ($t$-SNE) analysis of the element embedding vectors ($16 \rightarrow 2$ dimensions) shows that the NNs automatically develop their own concept of chemical similarity.}
	\label{fig:TSNE}
\end{figure}

Fig.~\ref{fig:AIvsDFT} provides an impression of the NN accuracy.
The prediction of the basal lattice parameters $a_0^\text{P}$ (not shown) and $a_0^\text{IL}$ proved to be straightforward,
whereas $c_0^\text{IL}$ turned out to be challenging.
This can be traced back to the sparse data available for vertically expanding materials [Fig.~\ref{fig:Structures}(c)]
and the only weak correlations of $c_0^\text{IL}$ with other observables [Fig.~\ref{fig:PSEs}(c)].
Here, AL significantly enhances the prediction accuracy as compared to a randomly chosen training set (Fig.~\ref{fig:AIvsDFT}).
As an example, boron at the $B$ site, combined with a post-transition-metal element at the $A$ site, tends to induce a large vertical expansion.
Already in the first iteration, these unconventional compounds are \textit{automatically} identified and included in the training set [Fig.~\ref{fig:AL}(c)].

AL-iterating towards $\sim 50\%$ training set size,
we already obtain a MAE $\sim 0.1$~eV for $E_f^{V_\text{O}}$ per vacancy (Fig.~\ref{fig:AIvsDFT}).
Relative to their range of $\sim 15$~eV, this corresponds to $<0.7\%$.
The heats of formation 
are predicted even more accurately,
reaching $25$~meV/atom (not shown), which is comparable to recent work on perovskites ($20$-$34$~meV/atom~\cite{YeOng:18}).
This reflects that $E_f^{V_\text{O}}$ is a fingerprint of the complex reaction and thus more demanding to predict.
For elpasolites, a heat of formation accuracy of $150$~meV/atom was obtained~\cite{Atom2VecPNAS:18}.
As a reference, the DFT accuracy can be considered as $\sim 0.1$~eV~\cite{XieGrossman:18, KirklinWolverton-DFTvsAI:15}.
A MAE of $0.2$~eV is achieved already around $35\%$.
A similar trend can be seen for the lattice parameters [Fig.~\ref{fig:AL}(b)].
In general, we observed that ensemble-averaged predictions of multiple NNs are more accurate than predictions by the individual NNs,
attaining $<1\%$ relative error for $\sim 35\%$ training set size.

Figure~\ref{fig:TSNE} explores the automatically generated NN element embedding vectors
by using stochastic neighbor embedding ($t$-SNE~\cite{tSNE:08}).
The nontrivial projection of a 16-dimensional space to two dimensions
reveals that the NNs develop a very unique understanding of the chemical similarity between the individual elements, mirroring the conventional picture of the periodic table.
This is even more compelling as the NNs are agnostic about concepts such as the atomic number or the group of a particular element.
In addition, we observed that this approach increases the accuracy as compared to directly passing the high-dimensional one-hot encoded element vectors to the densely connected hidden layers.

The AL algorithm can be stopped when the desired accuracy is reached [Fig.~\ref{fig:AL}(b)],
establishing the latter as a systematic control parameter.
Moreover, only the autonomously selected materials need to be calculated \textit{ab initio} in each iteration.
These aspects lead to a substantial gain in performance and energy efficiency as compared to conventional high-throughput calculations.
The presented methodology can be straightforwardly generalized to efficiently predict and enhance a broad scope of observables,
e.g., the thermoelectric performance~\cite{Geisler-LNOSTO:17, GeislerPentcheva-LNOLAO:18, TE-Fitness-Function-XingSingh:17},
across a large variety of interesting materials classes.


\begin{acknowledgments}

\textit{Acknowledgments.}
We thank Prof.~Dr.~Rossitza Pentcheva (University of Duisburg-Essen), Prof.~Dr.~Miguel Marques (University of Halle-Wittenberg), and Prof.~Dr.~Alexander Ecker (University of Göttingen) for helpful discussions.
B.G.\ acknowledges financial support 
provided by an Award in the Excellent Early Career Researchers Funding Competition of the University of Duisburg-Essen
and a Grant by the Department of Physics of the University of Duisburg-Essen.

\end{acknowledgments}


\begin{thebibliography}{60}%
\makeatletter
\providecommand \@ifxundefined [1]{%
 \@ifx{#1\undefined}
}%
\providecommand \@ifnum [1]{%
 \ifnum #1\expandafter \@firstoftwo
 \else \expandafter \@secondoftwo
 \fi
}%
\providecommand \@ifx [1]{%
 \ifx #1\expandafter \@firstoftwo
 \else \expandafter \@secondoftwo
 \fi
}%
\providecommand \natexlab [1]{#1}%
\providecommand \enquote  [1]{``#1''}%
\providecommand \bibnamefont  [1]{#1}%
\providecommand \bibfnamefont [1]{#1}%
\providecommand \citenamefont [1]{#1}%
\providecommand \href@noop [0]{\@secondoftwo}%
\providecommand \href [0]{\begingroup \@sanitize@url \@href}%
\providecommand \@href[1]{\@@startlink{#1}\@@href}%
\providecommand \@@href[1]{\endgroup#1\@@endlink}%
\providecommand \@sanitize@url [0]{\catcode `\\12\catcode `\$12\catcode
  `\&12\catcode `\#12\catcode `\^12\catcode `\_12\catcode `\%12\relax}%
\providecommand \@@startlink[1]{}%
\providecommand \@@endlink[0]{}%
\providecommand \url  [0]{\begingroup\@sanitize@url \@url }%
\providecommand \@url [1]{\endgroup\@href {#1}{\urlprefix }}%
\providecommand \urlprefix  [0]{URL }%
\providecommand \Eprint [0]{\href }%
\providecommand \doibase [0]{http://dx.doi.org/}%
\providecommand \selectlanguage [0]{\@gobble}%
\providecommand \bibinfo  [0]{\@secondoftwo}%
\providecommand \bibfield  [0]{\@secondoftwo}%
\providecommand \translation [1]{[#1]}%
\providecommand \BibitemOpen [0]{}%
\providecommand \bibitemStop [0]{}%
\providecommand \bibitemNoStop [0]{.\EOS\space}%
\providecommand \EOS [0]{\spacefactor3000\relax}%
\providecommand \BibitemShut  [1]{\csname bibitem#1\endcsname}%
\let\auto@bib@innerbib\@empty
\bibitem [{\citenamefont {Butler}\ \emph {et~al.}(2018)\citenamefont {Butler},
  \citenamefont {Davies}, \citenamefont {Cartwright}, \citenamefont {Isayev},\
  and\ \citenamefont {Walsh}}]{ButlerWalsh:18}%
  \BibitemOpen
  \bibfield  {author} {\bibinfo {author} {\bibfnamefont {K.~T.}\ \bibnamefont
  {Butler}}, \bibinfo {author} {\bibfnamefont {D.~W.}\ \bibnamefont {Davies}},
  \bibinfo {author} {\bibfnamefont {H.}~\bibnamefont {Cartwright}}, \bibinfo
  {author} {\bibfnamefont {O.}~\bibnamefont {Isayev}}, \ and\ \bibinfo {author}
  {\bibfnamefont {A.}~\bibnamefont {Walsh}},\ }\href {\doibase
  10.1038/s41586-018-0337-2} {\bibfield  {journal} {\bibinfo  {journal}
  {Nature}\ }\textbf {\bibinfo {volume} {559}},\ \bibinfo {pages} {547}
  (\bibinfo {year} {2018})}\BibitemShut {NoStop}%
\bibitem [{\citenamefont {Schmidt}\ \emph {et~al.}(2019)\citenamefont
  {Schmidt}, \citenamefont {Marques}, \citenamefont {Botti},\ and\
  \citenamefont {Marques}}]{SchmidtMarques:19}%
  \BibitemOpen
  \bibfield  {author} {\bibinfo {author} {\bibfnamefont {J.}~\bibnamefont
  {Schmidt}}, \bibinfo {author} {\bibfnamefont {M.~R.~G.}\ \bibnamefont
  {Marques}}, \bibinfo {author} {\bibfnamefont {S.}~\bibnamefont {Botti}}, \
  and\ \bibinfo {author} {\bibfnamefont {M.~A.~L.}\ \bibnamefont {Marques}},\
  }\href {\doibase 10.1038/s41524-019-0221-0} {\bibfield  {journal} {\bibinfo
  {journal} {npj Comput. Mater.}\ }\textbf {\bibinfo {volume} {5}},\ \bibinfo
  {pages} {83} (\bibinfo {year} {2019})}\BibitemShut {NoStop}%
\bibitem [{\citenamefont {Wang}\ \emph {et~al.}(2020)\citenamefont {Wang},
  \citenamefont {Murdock}, \citenamefont {Kauwe}, \citenamefont {Oliynyk},
  \citenamefont {Gurlo}, \citenamefont {Brgoch}, \citenamefont {Persson},\ and\
  \citenamefont {Sparks}}]{WangSparks:20}%
  \BibitemOpen
  \bibfield  {author} {\bibinfo {author} {\bibfnamefont {A.~Y.-T.}\
  \bibnamefont {Wang}}, \bibinfo {author} {\bibfnamefont {R.~J.}\ \bibnamefont
  {Murdock}}, \bibinfo {author} {\bibfnamefont {S.~K.}\ \bibnamefont {Kauwe}},
  \bibinfo {author} {\bibfnamefont {A.~O.}\ \bibnamefont {Oliynyk}}, \bibinfo
  {author} {\bibfnamefont {A.}~\bibnamefont {Gurlo}}, \bibinfo {author}
  {\bibfnamefont {J.}~\bibnamefont {Brgoch}}, \bibinfo {author} {\bibfnamefont
  {K.~A.}\ \bibnamefont {Persson}}, \ and\ \bibinfo {author} {\bibfnamefont
  {T.~D.}\ \bibnamefont {Sparks}},\ }\href {\doibase
  10.1021/acs.chemmater.0c01907} {\bibfield  {journal} {\bibinfo  {journal}
  {Chem. Mat.}\ }\textbf {\bibinfo {volume} {32}},\ \bibinfo {pages} {4954}
  (\bibinfo {year} {2020})}\BibitemShut {NoStop}%
\bibitem [{\citenamefont {Ward}\ \emph {et~al.}(2016)\citenamefont {Ward},
  \citenamefont {Agrawal}, \citenamefont {Choudhary},\ and\ \citenamefont
  {Wolverton}}]{WardWolverton:16}%
  \BibitemOpen
  \bibfield  {author} {\bibinfo {author} {\bibfnamefont {L.}~\bibnamefont
  {Ward}}, \bibinfo {author} {\bibfnamefont {A.}~\bibnamefont {Agrawal}},
  \bibinfo {author} {\bibfnamefont {A.}~\bibnamefont {Choudhary}}, \ and\
  \bibinfo {author} {\bibfnamefont {C.}~\bibnamefont {Wolverton}},\ }\href
  {\doibase 10.1038/npjcompumats.2016.28} {\bibfield  {journal} {\bibinfo
  {journal} {npj Comput. Mater.}\ }\textbf {\bibinfo {volume} {2}},\ \bibinfo
  {pages} {16028} (\bibinfo {year} {2016})}\BibitemShut {NoStop}%
\bibitem [{\citenamefont {Faber}\ \emph {et~al.}(2016)\citenamefont {Faber},
  \citenamefont {Lindmaa}, \citenamefont {von Lilienfeld},\ and\ \citenamefont
  {Armiento}}]{FaberLilienfeldArmiento-Elpasolite:16}%
  \BibitemOpen
  \bibfield  {author} {\bibinfo {author} {\bibfnamefont {F.~A.}\ \bibnamefont
  {Faber}}, \bibinfo {author} {\bibfnamefont {A.}~\bibnamefont {Lindmaa}},
  \bibinfo {author} {\bibfnamefont {O.~A.}\ \bibnamefont {von Lilienfeld}}, \
  and\ \bibinfo {author} {\bibfnamefont {R.}~\bibnamefont {Armiento}},\ }\href
  {\doibase 10.1103/PhysRevLett.117.135502} {\bibfield  {journal} {\bibinfo
  {journal} {Phys. Rev. Lett.}\ }\textbf {\bibinfo {volume} {117}},\ \bibinfo
  {pages} {135502} (\bibinfo {year} {2016})}\BibitemShut {NoStop}%
\bibitem [{\citenamefont {Ghiringhelli}\ \emph {et~al.}(2015)\citenamefont
  {Ghiringhelli}, \citenamefont {Vybiral}, \citenamefont {Levchenko},
  \citenamefont {Draxl},\ and\ \citenamefont
  {Scheffler}}]{Ghiringhelli-Descriptors:15}%
  \BibitemOpen
  \bibfield  {author} {\bibinfo {author} {\bibfnamefont {L.~M.}\ \bibnamefont
  {Ghiringhelli}}, \bibinfo {author} {\bibfnamefont {J.}~\bibnamefont
  {Vybiral}}, \bibinfo {author} {\bibfnamefont {S.~V.}\ \bibnamefont
  {Levchenko}}, \bibinfo {author} {\bibfnamefont {C.}~\bibnamefont {Draxl}}, \
  and\ \bibinfo {author} {\bibfnamefont {M.}~\bibnamefont {Scheffler}},\ }\href
  {\doibase 10.1103/PhysRevLett.114.105503} {\bibfield  {journal} {\bibinfo
  {journal} {Phys. Rev. Lett.}\ }\textbf {\bibinfo {volume} {114}},\ \bibinfo
  {pages} {105503} (\bibinfo {year} {2015})}\BibitemShut {NoStop}%
\bibitem [{\citenamefont {Bartel}\ \emph {et~al.}(2019)\citenamefont {Bartel},
  \citenamefont {Sutton}, \citenamefont {Goldsmith}, \citenamefont {Ouyang},
  \citenamefont {Musgrave}, \citenamefont {Ghiringhelli},\ and\ \citenamefont
  {Scheffler}}]{Bartel-ToleranceFactor:19}%
  \BibitemOpen
  \bibfield  {author} {\bibinfo {author} {\bibfnamefont {C.~J.}\ \bibnamefont
  {Bartel}}, \bibinfo {author} {\bibfnamefont {C.}~\bibnamefont {Sutton}},
  \bibinfo {author} {\bibfnamefont {B.~R.}\ \bibnamefont {Goldsmith}}, \bibinfo
  {author} {\bibfnamefont {R.}~\bibnamefont {Ouyang}}, \bibinfo {author}
  {\bibfnamefont {C.~B.}\ \bibnamefont {Musgrave}}, \bibinfo {author}
  {\bibfnamefont {L.~M.}\ \bibnamefont {Ghiringhelli}}, \ and\ \bibinfo
  {author} {\bibfnamefont {M.}~\bibnamefont {Scheffler}},\ }\href {\doibase
  10.1126/sciadv.aav0693} {\bibfield  {journal} {\bibinfo  {journal} {Science
  Advances}\ }\textbf {\bibinfo {volume} {5}} (\bibinfo {year} {2019}),\
  10.1126/sciadv.aav0693}\BibitemShut {NoStop}%
\bibitem [{\citenamefont {Schmidt}\ \emph {et~al.}(2017)\citenamefont
  {Schmidt}, \citenamefont {Shi}, \citenamefont {Borlido}, \citenamefont
  {Chen}, \citenamefont {Botti},\ and\ \citenamefont
  {Marques}}]{SchmidtMarques:17}%
  \BibitemOpen
  \bibfield  {author} {\bibinfo {author} {\bibfnamefont {J.}~\bibnamefont
  {Schmidt}}, \bibinfo {author} {\bibfnamefont {J.}~\bibnamefont {Shi}},
  \bibinfo {author} {\bibfnamefont {P.}~\bibnamefont {Borlido}}, \bibinfo
  {author} {\bibfnamefont {L.}~\bibnamefont {Chen}}, \bibinfo {author}
  {\bibfnamefont {S.}~\bibnamefont {Botti}}, \ and\ \bibinfo {author}
  {\bibfnamefont {M.~A.~L.}\ \bibnamefont {Marques}},\ }\href {\doibase
  10.1021/acs.chemmater.7b00156} {\bibfield  {journal} {\bibinfo  {journal}
  {Chem. Mat.}\ }\textbf {\bibinfo {volume} {29}},\ \bibinfo {pages} {5090}
  (\bibinfo {year} {2017})}\BibitemShut {NoStop}%
\bibitem [{\citenamefont {Belviso}\ \emph {et~al.}(2019)\citenamefont
  {Belviso}, \citenamefont {Claerbout}, \citenamefont {Comas-Vives},
  \citenamefont {Dalal}, \citenamefont {Fan}, \citenamefont {Filippetti},
  \citenamefont {Fiorentini}, \citenamefont {Foppa}, \citenamefont {Franchini},
  \citenamefont {Geisler}, \citenamefont {Ghiringhelli}, \citenamefont
  {Gro{\ss}}, \citenamefont {Hu}, \citenamefont {{\'I}{\~n}iguez},
  \citenamefont {Kauwe}, \citenamefont {Musfeldt}, \citenamefont {Nicolini},
  \citenamefont {Pentcheva}, \citenamefont {Polcar}, \citenamefont {Ren},
  \citenamefont {Ricci}, \citenamefont {Ricci}, \citenamefont {Sen},
  \citenamefont {Skelton}, \citenamefont {Sparks}, \citenamefont {Stroppa},
  \citenamefont {Urru}, \citenamefont {Vandichel}, \citenamefont {Vavassori},
  \citenamefont {Wu}, \citenamefont {Yang}, \citenamefont {Zhao}, \citenamefont
  {Puggioni}, \citenamefont {Cortese},\ and\ \citenamefont
  {Cammarata}}]{Viewpoint:19}%
  \BibitemOpen
  \bibfield  {author} {\bibinfo {author} {\bibfnamefont {F.}~\bibnamefont
  {Belviso}}, \bibinfo {author} {\bibfnamefont {V.~E.~P.}\ \bibnamefont
  {Claerbout}}, \bibinfo {author} {\bibfnamefont {A.}~\bibnamefont
  {Comas-Vives}}, \bibinfo {author} {\bibfnamefont {N.~S.}\ \bibnamefont
  {Dalal}}, \bibinfo {author} {\bibfnamefont {F.-R.}\ \bibnamefont {Fan}},
  \bibinfo {author} {\bibfnamefont {A.}~\bibnamefont {Filippetti}}, \bibinfo
  {author} {\bibfnamefont {V.}~\bibnamefont {Fiorentini}}, \bibinfo {author}
  {\bibfnamefont {L.}~\bibnamefont {Foppa}}, \bibinfo {author} {\bibfnamefont
  {C.}~\bibnamefont {Franchini}}, \bibinfo {author} {\bibfnamefont
  {B.}~\bibnamefont {Geisler}}, \bibinfo {author} {\bibfnamefont {L.~M.}\
  \bibnamefont {Ghiringhelli}}, \bibinfo {author} {\bibfnamefont
  {A.}~\bibnamefont {Gro{\ss}}}, \bibinfo {author} {\bibfnamefont
  {S.}~\bibnamefont {Hu}}, \bibinfo {author} {\bibfnamefont {J.}~\bibnamefont
  {{\'I}{\~n}iguez}}, \bibinfo {author} {\bibfnamefont {S.~K.}\ \bibnamefont
  {Kauwe}}, \bibinfo {author} {\bibfnamefont {J.~L.}\ \bibnamefont {Musfeldt}},
  \bibinfo {author} {\bibfnamefont {P.}~\bibnamefont {Nicolini}}, \bibinfo
  {author} {\bibfnamefont {R.}~\bibnamefont {Pentcheva}}, \bibinfo {author}
  {\bibfnamefont {T.}~\bibnamefont {Polcar}}, \bibinfo {author} {\bibfnamefont
  {W.}~\bibnamefont {Ren}}, \bibinfo {author} {\bibfnamefont {F.}~\bibnamefont
  {Ricci}}, \bibinfo {author} {\bibfnamefont {F.}~\bibnamefont {Ricci}},
  \bibinfo {author} {\bibfnamefont {H.~S.}\ \bibnamefont {Sen}}, \bibinfo
  {author} {\bibfnamefont {J.~M.}\ \bibnamefont {Skelton}}, \bibinfo {author}
  {\bibfnamefont {T.~D.}\ \bibnamefont {Sparks}}, \bibinfo {author}
  {\bibfnamefont {A.}~\bibnamefont {Stroppa}}, \bibinfo {author} {\bibfnamefont
  {A.}~\bibnamefont {Urru}}, \bibinfo {author} {\bibfnamefont {M.}~\bibnamefont
  {Vandichel}}, \bibinfo {author} {\bibfnamefont {P.}~\bibnamefont
  {Vavassori}}, \bibinfo {author} {\bibfnamefont {H.}~\bibnamefont {Wu}},
  \bibinfo {author} {\bibfnamefont {K.}~\bibnamefont {Yang}}, \bibinfo {author}
  {\bibfnamefont {H.~J.}\ \bibnamefont {Zhao}}, \bibinfo {author}
  {\bibfnamefont {D.}~\bibnamefont {Puggioni}}, \bibinfo {author}
  {\bibfnamefont {R.}~\bibnamefont {Cortese}}, \ and\ \bibinfo {author}
  {\bibfnamefont {A.}~\bibnamefont {Cammarata}},\ }\href {\doibase
  10.1021/acs.inorgchem.9b01785} {\bibfield  {journal} {\bibinfo  {journal}
  {Inorg. Chem.}\ }\textbf {\bibinfo {volume} {58}},\ \bibinfo {pages} {14939}
  (\bibinfo {year} {2019})}\BibitemShut {NoStop}%
\bibitem [{\citenamefont {Lopez-Bezanilla}\ and\ \citenamefont
  {Littlewood}(2020)}]{LopezLittlewood:20}%
  \BibitemOpen
  \bibfield  {author} {\bibinfo {author} {\bibfnamefont {A.}~\bibnamefont
  {Lopez-Bezanilla}}\ and\ \bibinfo {author} {\bibfnamefont {P.~B.}\
  \bibnamefont {Littlewood}},\ }\href {\doibase 10.1557/mrc.2020.2} {\bibfield
  {journal} {\bibinfo  {journal} {MRS Commun.}\ }\textbf {\bibinfo {volume}
  {10}},\ \bibinfo {pages} {1} (\bibinfo {year} {2020})}\BibitemShut {NoStop}%
\bibitem [{\citenamefont {Xie}\ and\ \citenamefont
  {Grossman}(2018)}]{XieGrossman:18}%
  \BibitemOpen
  \bibfield  {author} {\bibinfo {author} {\bibfnamefont {T.}~\bibnamefont
  {Xie}}\ and\ \bibinfo {author} {\bibfnamefont {J.~C.}\ \bibnamefont
  {Grossman}},\ }\href {\doibase 10.1103/PhysRevLett.120.145301} {\bibfield
  {journal} {\bibinfo  {journal} {Phys. Rev. Lett.}\ }\textbf {\bibinfo
  {volume} {120}},\ \bibinfo {pages} {145301} (\bibinfo {year}
  {2018})}\BibitemShut {NoStop}%
\bibitem [{\citenamefont {Ye}\ \emph {et~al.}(2018)\citenamefont {Ye},
  \citenamefont {Chen}, \citenamefont {Wang}, \citenamefont {Chu},\ and\
  \citenamefont {Ong}}]{YeOng:18}%
  \BibitemOpen
  \bibfield  {author} {\bibinfo {author} {\bibfnamefont {W.}~\bibnamefont
  {Ye}}, \bibinfo {author} {\bibfnamefont {C.}~\bibnamefont {Chen}}, \bibinfo
  {author} {\bibfnamefont {Z.}~\bibnamefont {Wang}}, \bibinfo {author}
  {\bibfnamefont {I.-H.}\ \bibnamefont {Chu}}, \ and\ \bibinfo {author}
  {\bibfnamefont {S.~P.}\ \bibnamefont {Ong}},\ }\href {\doibase
  10.1038/s41467-018-06322-x} {\bibfield  {journal} {\bibinfo  {journal} {Nat.
  Commun.}\ }\textbf {\bibinfo {volume} {9}},\ \bibinfo {pages} {3800}
  (\bibinfo {year} {2018})}\BibitemShut {NoStop}%
\bibitem [{\citenamefont {Jha}\ \emph {et~al.}(2019)\citenamefont {Jha},
  \citenamefont {Choudhary}, \citenamefont {Tavazza}, \citenamefont {Liao},
  \citenamefont {Choudhary}, \citenamefont {Campbell},\ and\ \citenamefont
  {Agrawal}}]{JhaAgrawal:19}%
  \BibitemOpen
  \bibfield  {author} {\bibinfo {author} {\bibfnamefont {D.}~\bibnamefont
  {Jha}}, \bibinfo {author} {\bibfnamefont {K.}~\bibnamefont {Choudhary}},
  \bibinfo {author} {\bibfnamefont {F.}~\bibnamefont {Tavazza}}, \bibinfo
  {author} {\bibfnamefont {W.-k.}\ \bibnamefont {Liao}}, \bibinfo {author}
  {\bibfnamefont {A.}~\bibnamefont {Choudhary}}, \bibinfo {author}
  {\bibfnamefont {C.}~\bibnamefont {Campbell}}, \ and\ \bibinfo {author}
  {\bibfnamefont {A.}~\bibnamefont {Agrawal}},\ }\href {\doibase
  10.1038/s41467-019-13297-w} {\bibfield  {journal} {\bibinfo  {journal} {Nat.
  Commun.}\ }\textbf {\bibinfo {volume} {10}},\ \bibinfo {pages} {5316}
  (\bibinfo {year} {2019})}\BibitemShut {NoStop}%
\bibitem [{\citenamefont {Agrawal}\ and\ \citenamefont
  {Choudhary}(2019)}]{AgrawalChoudhary:19}%
  \BibitemOpen
  \bibfield  {author} {\bibinfo {author} {\bibfnamefont {A.}~\bibnamefont
  {Agrawal}}\ and\ \bibinfo {author} {\bibfnamefont {A.}~\bibnamefont
  {Choudhary}},\ }\href {\doibase 10.1557/mrc.2019.73} {\bibfield  {journal}
  {\bibinfo  {journal} {MRS Commun.}\ }\textbf {\bibinfo {volume} {9}},\
  \bibinfo {pages} {779} (\bibinfo {year} {2019})}\BibitemShut {NoStop}%
\bibitem [{\citenamefont {Zhang}\ \emph {et~al.}(2018)\citenamefont {Zhang},
  \citenamefont {Han}, \citenamefont {Wang}, \citenamefont {Car},\ and\
  \citenamefont {E}}]{DL-MD-Car:18}%
  \BibitemOpen
  \bibfield  {author} {\bibinfo {author} {\bibfnamefont {L.}~\bibnamefont
  {Zhang}}, \bibinfo {author} {\bibfnamefont {J.}~\bibnamefont {Han}}, \bibinfo
  {author} {\bibfnamefont {H.}~\bibnamefont {Wang}}, \bibinfo {author}
  {\bibfnamefont {R.}~\bibnamefont {Car}}, \ and\ \bibinfo {author}
  {\bibfnamefont {W.}~\bibnamefont {E}},\ }\href {\doibase
  10.1103/PhysRevLett.120.143001} {\bibfield  {journal} {\bibinfo  {journal}
  {Phys. Rev. Lett.}\ }\textbf {\bibinfo {volume} {120}},\ \bibinfo {pages}
  {143001} (\bibinfo {year} {2018})}\BibitemShut {NoStop}%
\bibitem [{\citenamefont {Tshitoyan}\ \emph {et~al.}(2019)\citenamefont
  {Tshitoyan}, \citenamefont {Dagdelen}, \citenamefont {Weston}, \citenamefont
  {Dunn}, \citenamefont {Rong}, \citenamefont {Kononova}, \citenamefont
  {Persson}, \citenamefont {Ceder},\ and\ \citenamefont
  {Jain}}]{TshitoyanJain-UnsupervisedWordEmbedding:19}%
  \BibitemOpen
  \bibfield  {author} {\bibinfo {author} {\bibfnamefont {V.}~\bibnamefont
  {Tshitoyan}}, \bibinfo {author} {\bibfnamefont {J.}~\bibnamefont {Dagdelen}},
  \bibinfo {author} {\bibfnamefont {L.}~\bibnamefont {Weston}}, \bibinfo
  {author} {\bibfnamefont {A.}~\bibnamefont {Dunn}}, \bibinfo {author}
  {\bibfnamefont {Z.}~\bibnamefont {Rong}}, \bibinfo {author} {\bibfnamefont
  {O.}~\bibnamefont {Kononova}}, \bibinfo {author} {\bibfnamefont {K.~A.}\
  \bibnamefont {Persson}}, \bibinfo {author} {\bibfnamefont {G.}~\bibnamefont
  {Ceder}}, \ and\ \bibinfo {author} {\bibfnamefont {A.}~\bibnamefont {Jain}},\
  }\href {\doibase 10.1038/s41586-019-1335-8} {\bibfield  {journal} {\bibinfo
  {journal} {Nature}\ }\textbf {\bibinfo {volume} {571}},\ \bibinfo {pages}
  {95} (\bibinfo {year} {2019})}\BibitemShut {NoStop}%
\bibitem [{\citenamefont {Zhou}\ \emph {et~al.}(2018)\citenamefont {Zhou},
  \citenamefont {Tang}, \citenamefont {Liu}, \citenamefont {Pan}, \citenamefont
  {Yan},\ and\ \citenamefont {Zhang}}]{Atom2VecPNAS:18}%
  \BibitemOpen
  \bibfield  {author} {\bibinfo {author} {\bibfnamefont {Q.}~\bibnamefont
  {Zhou}}, \bibinfo {author} {\bibfnamefont {P.}~\bibnamefont {Tang}}, \bibinfo
  {author} {\bibfnamefont {S.}~\bibnamefont {Liu}}, \bibinfo {author}
  {\bibfnamefont {J.}~\bibnamefont {Pan}}, \bibinfo {author} {\bibfnamefont
  {Q.}~\bibnamefont {Yan}}, \ and\ \bibinfo {author} {\bibfnamefont {S.-C.}\
  \bibnamefont {Zhang}},\ }\href {\doibase 10.1073/pnas.1801181115} {\bibfield
  {journal} {\bibinfo  {journal} {PNAS}\ }\textbf {\bibinfo {volume} {115}},\
  \bibinfo {pages} {E6411} (\bibinfo {year} {2018})}\BibitemShut {NoStop}%
\bibitem [{\citenamefont {Jha}\ \emph {et~al.}(2018)\citenamefont {Jha},
  \citenamefont {Ward}, \citenamefont {Paul}, \citenamefont {Liao},
  \citenamefont {Choudhary}, \citenamefont {Wolverton},\ and\ \citenamefont
  {Agrawal}}]{JhaWolvertonAgrawal:18}%
  \BibitemOpen
  \bibfield  {author} {\bibinfo {author} {\bibfnamefont {D.}~\bibnamefont
  {Jha}}, \bibinfo {author} {\bibfnamefont {L.}~\bibnamefont {Ward}}, \bibinfo
  {author} {\bibfnamefont {A.}~\bibnamefont {Paul}}, \bibinfo {author}
  {\bibfnamefont {W.-k.}\ \bibnamefont {Liao}}, \bibinfo {author}
  {\bibfnamefont {A.}~\bibnamefont {Choudhary}}, \bibinfo {author}
  {\bibfnamefont {C.}~\bibnamefont {Wolverton}}, \ and\ \bibinfo {author}
  {\bibfnamefont {A.}~\bibnamefont {Agrawal}},\ }\href {\doibase
  10.1038/s41598-018-35934-y} {\bibfield  {journal} {\bibinfo  {journal} {Sci.
  Rep.}\ }\textbf {\bibinfo {volume} {8}},\ \bibinfo {pages} {17593} (\bibinfo
  {year} {2018})}\BibitemShut {NoStop}%
\bibitem [{\citenamefont {Jain}\ \emph {et~al.}(2013)\citenamefont {Jain},
  \citenamefont {Ong}, \citenamefont {Hautier}, \citenamefont {Chen},
  \citenamefont {Richards}, \citenamefont {Dacek}, \citenamefont {Cholia},
  \citenamefont {Gunter}, \citenamefont {Skinner}, \citenamefont {Ceder},\ and\
  \citenamefont {Persson}}]{MatProj:13}%
  \BibitemOpen
  \bibfield  {author} {\bibinfo {author} {\bibfnamefont {A.}~\bibnamefont
  {Jain}}, \bibinfo {author} {\bibfnamefont {S.~P.}\ \bibnamefont {Ong}},
  \bibinfo {author} {\bibfnamefont {G.}~\bibnamefont {Hautier}}, \bibinfo
  {author} {\bibfnamefont {W.}~\bibnamefont {Chen}}, \bibinfo {author}
  {\bibfnamefont {W.~D.}\ \bibnamefont {Richards}}, \bibinfo {author}
  {\bibfnamefont {S.}~\bibnamefont {Dacek}}, \bibinfo {author} {\bibfnamefont
  {S.}~\bibnamefont {Cholia}}, \bibinfo {author} {\bibfnamefont
  {D.}~\bibnamefont {Gunter}}, \bibinfo {author} {\bibfnamefont
  {D.}~\bibnamefont {Skinner}}, \bibinfo {author} {\bibfnamefont
  {G.}~\bibnamefont {Ceder}}, \ and\ \bibinfo {author} {\bibfnamefont {K.~A.}\
  \bibnamefont {Persson}},\ }\href {\doibase 10.1063/1.4812323} {\bibfield
  {journal} {\bibinfo  {journal} {APL Materials}\ }\textbf {\bibinfo {volume}
  {1}},\ \bibinfo {pages} {011002} (\bibinfo {year} {2013})}\BibitemShut
  {NoStop}%
\bibitem [{\citenamefont {Saal}\ \emph {et~al.}(2013)\citenamefont {Saal},
  \citenamefont {Kirklin}, \citenamefont {Aykol}, \citenamefont {Meredig},\
  and\ \citenamefont {Wolverton}}]{OQMD:13}%
  \BibitemOpen
  \bibfield  {author} {\bibinfo {author} {\bibfnamefont {J.~E.}\ \bibnamefont
  {Saal}}, \bibinfo {author} {\bibfnamefont {S.}~\bibnamefont {Kirklin}},
  \bibinfo {author} {\bibfnamefont {M.}~\bibnamefont {Aykol}}, \bibinfo
  {author} {\bibfnamefont {B.}~\bibnamefont {Meredig}}, \ and\ \bibinfo
  {author} {\bibfnamefont {C.}~\bibnamefont {Wolverton}},\ }\href {\doibase
  10.1007/s11837-013-0755-4} {\bibfield  {journal} {\bibinfo  {journal} {JOM}\
  }\textbf {\bibinfo {volume} {65}},\ \bibinfo {pages} {1501} (\bibinfo {year}
  {2013})}\BibitemShut {NoStop}%
\bibitem [{\citenamefont {Curtarolo}\ \emph {et~al.}(2012)\citenamefont
  {Curtarolo}, \citenamefont {Setyawan}, \citenamefont {Hart}, \citenamefont
  {Jahnatek}, \citenamefont {Chepulskii}, \citenamefont {Taylor}, \citenamefont
  {Wang}, \citenamefont {Xue}, \citenamefont {Yang}, \citenamefont {Levy},
  \citenamefont {Mehl}, \citenamefont {Stokes}, \citenamefont {Demchenko},\
  and\ \citenamefont {Morgan}}]{AFLOW:12}%
  \BibitemOpen
  \bibfield  {author} {\bibinfo {author} {\bibfnamefont {S.}~\bibnamefont
  {Curtarolo}}, \bibinfo {author} {\bibfnamefont {W.}~\bibnamefont {Setyawan}},
  \bibinfo {author} {\bibfnamefont {G.~L.}\ \bibnamefont {Hart}}, \bibinfo
  {author} {\bibfnamefont {M.}~\bibnamefont {Jahnatek}}, \bibinfo {author}
  {\bibfnamefont {R.~V.}\ \bibnamefont {Chepulskii}}, \bibinfo {author}
  {\bibfnamefont {R.~H.}\ \bibnamefont {Taylor}}, \bibinfo {author}
  {\bibfnamefont {S.}~\bibnamefont {Wang}}, \bibinfo {author} {\bibfnamefont
  {J.}~\bibnamefont {Xue}}, \bibinfo {author} {\bibfnamefont {K.}~\bibnamefont
  {Yang}}, \bibinfo {author} {\bibfnamefont {O.}~\bibnamefont {Levy}}, \bibinfo
  {author} {\bibfnamefont {M.~J.}\ \bibnamefont {Mehl}}, \bibinfo {author}
  {\bibfnamefont {H.~T.}\ \bibnamefont {Stokes}}, \bibinfo {author}
  {\bibfnamefont {D.~O.}\ \bibnamefont {Demchenko}}, \ and\ \bibinfo {author}
  {\bibfnamefont {D.}~\bibnamefont {Morgan}},\ }\href {\doibase
  https://doi.org/10.1016/j.commatsci.2012.02.005} {\bibfield  {journal}
  {\bibinfo  {journal} {Comp. Mat. Sci.}\ }\textbf {\bibinfo {volume} {58}},\
  \bibinfo {pages} {218} (\bibinfo {year} {2012})}\BibitemShut {NoStop}%
\bibitem [{\citenamefont {Draxl}\ and\ \citenamefont
  {Scheffler}(2018)}]{NOMAD:18}%
  \BibitemOpen
  \bibfield  {author} {\bibinfo {author} {\bibfnamefont {C.}~\bibnamefont
  {Draxl}}\ and\ \bibinfo {author} {\bibfnamefont {M.}~\bibnamefont
  {Scheffler}},\ }\href {\doibase 10.1557/mrs.2018.208} {\bibfield  {journal}
  {\bibinfo  {journal} {MRS Bulletin}\ }\textbf {\bibinfo {volume} {43}},\
  \bibinfo {pages} {676} (\bibinfo {year} {2018})}\BibitemShut {NoStop}%
\bibitem [{\citenamefont {Li}\ \emph {et~al.}(2019)\citenamefont {Li},
  \citenamefont {Lee}, \citenamefont {Wang}, \citenamefont {Osada},
  \citenamefont {Crossley}, \citenamefont {Lee}, \citenamefont {Cui},
  \citenamefont {Hikita},\ and\ \citenamefont
  {Hwang}}]{Li-Supercond-Inf-NNO-STO:19}%
  \BibitemOpen
  \bibfield  {author} {\bibinfo {author} {\bibfnamefont {D.}~\bibnamefont
  {Li}}, \bibinfo {author} {\bibfnamefont {K.}~\bibnamefont {Lee}}, \bibinfo
  {author} {\bibfnamefont {B.~Y.}\ \bibnamefont {Wang}}, \bibinfo {author}
  {\bibfnamefont {M.}~\bibnamefont {Osada}}, \bibinfo {author} {\bibfnamefont
  {S.}~\bibnamefont {Crossley}}, \bibinfo {author} {\bibfnamefont {H.~R.}\
  \bibnamefont {Lee}}, \bibinfo {author} {\bibfnamefont {Y.}~\bibnamefont
  {Cui}}, \bibinfo {author} {\bibfnamefont {Y.}~\bibnamefont {Hikita}}, \ and\
  \bibinfo {author} {\bibfnamefont {H.~Y.}\ \bibnamefont {Hwang}},\ }\href
  {\doibase 10.1038/s41586-019-1496-5} {\bibfield  {journal} {\bibinfo
  {journal} {Nature}\ }\textbf {\bibinfo {volume} {572}},\ \bibinfo {pages}
  {624} (\bibinfo {year} {2019})}\BibitemShut {NoStop}%
\bibitem [{\citenamefont {Nomura}\ \emph {et~al.}(2019)\citenamefont {Nomura},
  \citenamefont {Hirayama}, \citenamefont {Tadano}, \citenamefont {Yoshimoto},
  \citenamefont {Nakamura},\ and\ \citenamefont {Arita}}]{Nomura-Inf-NNO:19}%
  \BibitemOpen
  \bibfield  {author} {\bibinfo {author} {\bibfnamefont {Y.}~\bibnamefont
  {Nomura}}, \bibinfo {author} {\bibfnamefont {M.}~\bibnamefont {Hirayama}},
  \bibinfo {author} {\bibfnamefont {T.}~\bibnamefont {Tadano}}, \bibinfo
  {author} {\bibfnamefont {Y.}~\bibnamefont {Yoshimoto}}, \bibinfo {author}
  {\bibfnamefont {K.}~\bibnamefont {Nakamura}}, \ and\ \bibinfo {author}
  {\bibfnamefont {R.}~\bibnamefont {Arita}},\ }\href {\doibase
  10.1103/PhysRevB.100.205138} {\bibfield  {journal} {\bibinfo  {journal}
  {Phys. Rev. B}\ }\textbf {\bibinfo {volume} {100}},\ \bibinfo {pages}
  {205138} (\bibinfo {year} {2019})}\BibitemShut {NoStop}%
\bibitem [{\citenamefont {Jiang}\ \emph {et~al.}(2019)\citenamefont {Jiang},
  \citenamefont {Si}, \citenamefont {Liao},\ and\ \citenamefont
  {Zhong}}]{JiangZhong-InfNickelates:19}%
  \BibitemOpen
  \bibfield  {author} {\bibinfo {author} {\bibfnamefont {P.}~\bibnamefont
  {Jiang}}, \bibinfo {author} {\bibfnamefont {L.}~\bibnamefont {Si}}, \bibinfo
  {author} {\bibfnamefont {Z.}~\bibnamefont {Liao}}, \ and\ \bibinfo {author}
  {\bibfnamefont {Z.}~\bibnamefont {Zhong}},\ }\href {\doibase
  10.1103/PhysRevB.100.201106} {\bibfield  {journal} {\bibinfo  {journal}
  {Phys. Rev. B}\ }\textbf {\bibinfo {volume} {100}},\ \bibinfo {pages}
  {201106} (\bibinfo {year} {2019})}\BibitemShut {NoStop}%
\bibitem [{\citenamefont {Sakakibara}\ \emph {et~al.}(2020)\citenamefont
  {Sakakibara}, \citenamefont {Usui}, \citenamefont {Suzuki}, \citenamefont
  {Kotani}, \citenamefont {Aoki},\ and\ \citenamefont
  {Kuroki}}]{Sakakibara:19}%
  \BibitemOpen
  \bibfield  {author} {\bibinfo {author} {\bibfnamefont {H.}~\bibnamefont
  {Sakakibara}}, \bibinfo {author} {\bibfnamefont {H.}~\bibnamefont {Usui}},
  \bibinfo {author} {\bibfnamefont {K.}~\bibnamefont {Suzuki}}, \bibinfo
  {author} {\bibfnamefont {T.}~\bibnamefont {Kotani}}, \bibinfo {author}
  {\bibfnamefont {H.}~\bibnamefont {Aoki}}, \ and\ \bibinfo {author}
  {\bibfnamefont {K.}~\bibnamefont {Kuroki}},\ }\href {\doibase
  10.1103/PhysRevLett.125.077003} {\bibfield  {journal} {\bibinfo  {journal}
  {Phys. Rev. Lett.}\ }\textbf {\bibinfo {volume} {125}},\ \bibinfo {pages}
  {077003} (\bibinfo {year} {2020})}\BibitemShut {NoStop}%
\bibitem [{\citenamefont {Jiang}\ \emph {et~al.}(2020)\citenamefont {Jiang},
  \citenamefont {Berciu},\ and\ \citenamefont
  {Sawatzky}}]{JiangBerciuSawatzky:19}%
  \BibitemOpen
  \bibfield  {author} {\bibinfo {author} {\bibfnamefont {M.}~\bibnamefont
  {Jiang}}, \bibinfo {author} {\bibfnamefont {M.}~\bibnamefont {Berciu}}, \
  and\ \bibinfo {author} {\bibfnamefont {G.~A.}\ \bibnamefont {Sawatzky}},\
  }\href {\doibase 10.1103/PhysRevLett.124.207004} {\bibfield  {journal}
  {\bibinfo  {journal} {Phys. Rev. Lett.}\ }\textbf {\bibinfo {volume} {124}},\
  \bibinfo {pages} {207004} (\bibinfo {year} {2020})}\BibitemShut {NoStop}%
\bibitem [{\citenamefont {Botana}\ and\ \citenamefont
  {Norman}(2020)}]{Botana-Inf-Nickelates:19}%
  \BibitemOpen
  \bibfield  {author} {\bibinfo {author} {\bibfnamefont {A.~S.}\ \bibnamefont
  {Botana}}\ and\ \bibinfo {author} {\bibfnamefont {M.~R.}\ \bibnamefont
  {Norman}},\ }\href {\doibase 10.1103/PhysRevX.10.011024} {\bibfield
  {journal} {\bibinfo  {journal} {Phys. Rev. X}\ }\textbf {\bibinfo {volume}
  {10}},\ \bibinfo {pages} {011024} (\bibinfo {year} {2020})}\BibitemShut
  {NoStop}%
\bibitem [{\citenamefont {Osada}\ \emph {et~al.}(2020)\citenamefont {Osada},
  \citenamefont {Wang}, \citenamefont {Goodge}, \citenamefont {Lee},
  \citenamefont {Yoon}, \citenamefont {Sakuma}, \citenamefont {Li},
  \citenamefont {Miura}, \citenamefont {Kourkoutis},\ and\ \citenamefont
  {Hwang}}]{Osada-PrNiO2-SC:20}%
  \BibitemOpen
  \bibfield  {author} {\bibinfo {author} {\bibfnamefont {M.}~\bibnamefont
  {Osada}}, \bibinfo {author} {\bibfnamefont {B.~Y.}\ \bibnamefont {Wang}},
  \bibinfo {author} {\bibfnamefont {B.~H.}\ \bibnamefont {Goodge}}, \bibinfo
  {author} {\bibfnamefont {K.}~\bibnamefont {Lee}}, \bibinfo {author}
  {\bibfnamefont {H.}~\bibnamefont {Yoon}}, \bibinfo {author} {\bibfnamefont
  {K.}~\bibnamefont {Sakuma}}, \bibinfo {author} {\bibfnamefont
  {D.}~\bibnamefont {Li}}, \bibinfo {author} {\bibfnamefont {M.}~\bibnamefont
  {Miura}}, \bibinfo {author} {\bibfnamefont {L.~F.}\ \bibnamefont
  {Kourkoutis}}, \ and\ \bibinfo {author} {\bibfnamefont {H.~Y.}\ \bibnamefont
  {Hwang}},\ }\href {\doibase 10.1021/acs.nanolett.0c01392} {\bibfield
  {journal} {\bibinfo  {journal} {Nano Lett.}\ }\textbf {\bibinfo {volume}
  {20}},\ \bibinfo {pages} {5735} (\bibinfo {year} {2020})}\BibitemShut
  {NoStop}%
\bibitem [{\citenamefont {Lechermann}(2020)}]{Lechermann-Inf:20}%
  \BibitemOpen
  \bibfield  {author} {\bibinfo {author} {\bibfnamefont {F.}~\bibnamefont
  {Lechermann}},\ }\href {\doibase 10.1103/PhysRevB.101.081110} {\bibfield
  {journal} {\bibinfo  {journal} {Phys. Rev. B}\ }\textbf {\bibinfo {volume}
  {101}},\ \bibinfo {pages} {081110} (\bibinfo {year} {2020})}\BibitemShut
  {NoStop}%
\bibitem [{\citenamefont {Wu}\ \emph {et~al.}(2020)\citenamefont {Wu},
  \citenamefont {Di~Sante}, \citenamefont {Schwemmer}, \citenamefont {Hanke},
  \citenamefont {Hwang}, \citenamefont {Raghu},\ and\ \citenamefont
  {Thomale}}]{NNO-SC-Thomale:20}%
  \BibitemOpen
  \bibfield  {author} {\bibinfo {author} {\bibfnamefont {X.}~\bibnamefont
  {Wu}}, \bibinfo {author} {\bibfnamefont {D.}~\bibnamefont {Di~Sante}},
  \bibinfo {author} {\bibfnamefont {T.}~\bibnamefont {Schwemmer}}, \bibinfo
  {author} {\bibfnamefont {W.}~\bibnamefont {Hanke}}, \bibinfo {author}
  {\bibfnamefont {H.~Y.}\ \bibnamefont {Hwang}}, \bibinfo {author}
  {\bibfnamefont {S.}~\bibnamefont {Raghu}}, \ and\ \bibinfo {author}
  {\bibfnamefont {R.}~\bibnamefont {Thomale}},\ }\href {\doibase
  10.1103/PhysRevB.101.060504} {\bibfield  {journal} {\bibinfo  {journal}
  {Phys. Rev. B}\ }\textbf {\bibinfo {volume} {101}},\ \bibinfo {pages}
  {060504} (\bibinfo {year} {2020})}\BibitemShut {NoStop}%
\bibitem [{\citenamefont {Hirayama}\ \emph {et~al.}(2020)\citenamefont
  {Hirayama}, \citenamefont {Tadano}, \citenamefont {Nomura},\ and\
  \citenamefont {Arita}}]{NNO-SelfDopingDesign-d9-Arita:20}%
  \BibitemOpen
  \bibfield  {author} {\bibinfo {author} {\bibfnamefont {M.}~\bibnamefont
  {Hirayama}}, \bibinfo {author} {\bibfnamefont {T.}~\bibnamefont {Tadano}},
  \bibinfo {author} {\bibfnamefont {Y.}~\bibnamefont {Nomura}}, \ and\ \bibinfo
  {author} {\bibfnamefont {R.}~\bibnamefont {Arita}},\ }\href {\doibase
  10.1103/PhysRevB.101.075107} {\bibfield  {journal} {\bibinfo  {journal}
  {Phys. Rev. B}\ }\textbf {\bibinfo {volume} {101}},\ \bibinfo {pages}
  {075107} (\bibinfo {year} {2020})}\BibitemShut {NoStop}%
\bibitem [{\citenamefont {Kitatani}\ \emph {et~al.}(2020)\citenamefont
  {Kitatani}, \citenamefont {Si}, \citenamefont {Janson}, \citenamefont
  {Arita}, \citenamefont {Zhong},\ and\ \citenamefont
  {Held}}]{Kitatani-AritaZhongHeld:20}%
  \BibitemOpen
  \bibfield  {author} {\bibinfo {author} {\bibfnamefont {M.}~\bibnamefont
  {Kitatani}}, \bibinfo {author} {\bibfnamefont {L.}~\bibnamefont {Si}},
  \bibinfo {author} {\bibfnamefont {O.}~\bibnamefont {Janson}}, \bibinfo
  {author} {\bibfnamefont {R.}~\bibnamefont {Arita}}, \bibinfo {author}
  {\bibfnamefont {Z.}~\bibnamefont {Zhong}}, \ and\ \bibinfo {author}
  {\bibfnamefont {K.}~\bibnamefont {Held}},\ }\href {\doibase
  10.1038/s41535-020-00260-y} {\bibfield  {journal} {\bibinfo  {journal} {npj
  Quantum Materials}\ }\textbf {\bibinfo {volume} {5}},\ \bibinfo {pages} {59}
  (\bibinfo {year} {2020})}\BibitemShut {NoStop}%
\bibitem [{\citenamefont {Geisler}\ and\ \citenamefont
  {Pentcheva}(2020{\natexlab{a}})}]{GeislerPentcheva-InfNNO:20}%
  \BibitemOpen
  \bibfield  {author} {\bibinfo {author} {\bibfnamefont {B.}~\bibnamefont
  {Geisler}}\ and\ \bibinfo {author} {\bibfnamefont {R.}~\bibnamefont
  {Pentcheva}},\ }\href {\doibase 10.1103/PhysRevB.102.020502} {\bibfield
  {journal} {\bibinfo  {journal} {Phys. Rev. B}\ }\textbf {\bibinfo {volume}
  {102}},\ \bibinfo {pages} {020502(R)} (\bibinfo {year}
  {2020}{\natexlab{a}})}\BibitemShut {NoStop}%
\bibitem [{\citenamefont {Li}\ \emph {et~al.}(2020)\citenamefont {Li},
  \citenamefont {Wang}, \citenamefont {Lee}, \citenamefont {Harvey},
  \citenamefont {Osada}, \citenamefont {Goodge}, \citenamefont {Kourkoutis},\
  and\ \citenamefont {Hwang}}]{Li-Supercond-Dome-Inf-NNO-STO:20}%
  \BibitemOpen
  \bibfield  {author} {\bibinfo {author} {\bibfnamefont {D.}~\bibnamefont
  {Li}}, \bibinfo {author} {\bibfnamefont {B.~Y.}\ \bibnamefont {Wang}},
  \bibinfo {author} {\bibfnamefont {K.}~\bibnamefont {Lee}}, \bibinfo {author}
  {\bibfnamefont {S.~P.}\ \bibnamefont {Harvey}}, \bibinfo {author}
  {\bibfnamefont {M.}~\bibnamefont {Osada}}, \bibinfo {author} {\bibfnamefont
  {B.~H.}\ \bibnamefont {Goodge}}, \bibinfo {author} {\bibfnamefont {L.~F.}\
  \bibnamefont {Kourkoutis}}, \ and\ \bibinfo {author} {\bibfnamefont {H.~Y.}\
  \bibnamefont {Hwang}},\ }\href {\doibase 10.1103/PhysRevLett.125.027001}
  {\bibfield  {journal} {\bibinfo  {journal} {Phys. Rev. Lett.}\ }\textbf
  {\bibinfo {volume} {125}},\ \bibinfo {pages} {027001} (\bibinfo {year}
  {2020})}\BibitemShut {NoStop}%
\bibitem [{\citenamefont {Choi}\ \emph {et~al.}(2020)\citenamefont {Choi},
  \citenamefont {Lee},\ and\ \citenamefont
  {Pickett}}]{Choi-Lee-Pickett-4fNNO:20}%
  \BibitemOpen
  \bibfield  {author} {\bibinfo {author} {\bibfnamefont {M.-Y.}\ \bibnamefont
  {Choi}}, \bibinfo {author} {\bibfnamefont {K.-W.}\ \bibnamefont {Lee}}, \
  and\ \bibinfo {author} {\bibfnamefont {W.~E.}\ \bibnamefont {Pickett}},\
  }\href {\doibase 10.1103/PhysRevB.101.020503} {\bibfield  {journal} {\bibinfo
   {journal} {Phys. Rev. B}\ }\textbf {\bibinfo {volume} {101}},\ \bibinfo
  {pages} {020503} (\bibinfo {year} {2020})}\BibitemShut {NoStop}%
\bibitem [{\citenamefont {Si}\ \emph {et~al.}(2020)\citenamefont {Si},
  \citenamefont {Xiao}, \citenamefont {Kaufmann}, \citenamefont {Tomczak},
  \citenamefont {Lu}, \citenamefont {Zhong},\ and\ \citenamefont
  {Held}}]{Si-Zhonh-Held:InfNNO-Hydrogen:20}%
  \BibitemOpen
  \bibfield  {author} {\bibinfo {author} {\bibfnamefont {L.}~\bibnamefont
  {Si}}, \bibinfo {author} {\bibfnamefont {W.}~\bibnamefont {Xiao}}, \bibinfo
  {author} {\bibfnamefont {J.}~\bibnamefont {Kaufmann}}, \bibinfo {author}
  {\bibfnamefont {J.~M.}\ \bibnamefont {Tomczak}}, \bibinfo {author}
  {\bibfnamefont {Y.}~\bibnamefont {Lu}}, \bibinfo {author} {\bibfnamefont
  {Z.}~\bibnamefont {Zhong}}, \ and\ \bibinfo {author} {\bibfnamefont
  {K.}~\bibnamefont {Held}},\ }\href {\doibase 10.1103/PhysRevLett.124.166402}
  {\bibfield  {journal} {\bibinfo  {journal} {Phys. Rev. Lett.}\ }\textbf
  {\bibinfo {volume} {124}},\ \bibinfo {pages} {166402} (\bibinfo {year}
  {2020})}\BibitemShut {NoStop}%
\bibitem [{\citenamefont {Geisler}\ and\ \citenamefont
  {Pentcheva}(2021)}]{GeislerPentcheva-NNOCCOSTO:21}%
  \BibitemOpen
  \bibfield  {author} {\bibinfo {author} {\bibfnamefont {B.}~\bibnamefont
  {Geisler}}\ and\ \bibinfo {author} {\bibfnamefont {R.}~\bibnamefont
  {Pentcheva}},\ }\href {\doibase 10.1103/PhysRevResearch.3.013261} {\bibfield
  {journal} {\bibinfo  {journal} {Phys. Rev. Research}\ }\textbf {\bibinfo
  {volume} {3}},\ \bibinfo {pages} {013261} (\bibinfo {year}
  {2021})}\BibitemShut {NoStop}%
\bibitem [{\citenamefont {Ortiz}\ \emph {et~al.}(2021)\citenamefont {Ortiz},
  \citenamefont {Mantadakis}, \citenamefont {Misják}, \citenamefont {Fürsich},
  \citenamefont {Schierle}, \citenamefont {Christiani}, \citenamefont
  {Logvenov}, \citenamefont {Kaiser}, \citenamefont {Keimer}, \citenamefont
  {Hansmann},\ and\ \citenamefont {Benckiser}}]{Ortiz-NNO:21}%
  \BibitemOpen
  \bibfield  {author} {\bibinfo {author} {\bibfnamefont {R.~A.}\ \bibnamefont
  {Ortiz}}, \bibinfo {author} {\bibfnamefont {D.~T.}\ \bibnamefont
  {Mantadakis}}, \bibinfo {author} {\bibfnamefont {F.}~\bibnamefont {Misják}},
  \bibinfo {author} {\bibfnamefont {K.}~\bibnamefont {Fürsich}}, \bibinfo
  {author} {\bibfnamefont {E.}~\bibnamefont {Schierle}}, \bibinfo {author}
  {\bibfnamefont {G.}~\bibnamefont {Christiani}}, \bibinfo {author}
  {\bibfnamefont {G.}~\bibnamefont {Logvenov}}, \bibinfo {author}
  {\bibfnamefont {U.}~\bibnamefont {Kaiser}}, \bibinfo {author} {\bibfnamefont
  {B.}~\bibnamefont {Keimer}}, \bibinfo {author} {\bibfnamefont
  {P.}~\bibnamefont {Hansmann}}, \ and\ \bibinfo {author} {\bibfnamefont
  {E.}~\bibnamefont {Benckiser}},\ }\href@noop {} {\enquote {\bibinfo {title}
  {A superlattice approach to doping infinite-layer nickelates},}\ } (\bibinfo
  {year} {2021}),\ \Eprint {http://arxiv.org/abs/2102.05621} {arXiv:2102.05621
  [cond-mat.supr-con]} \BibitemShut {NoStop}%
\bibitem [{\citenamefont {Mikolov}\ \emph {et~al.}(2013)\citenamefont
  {Mikolov}, \citenamefont {Chen}, \citenamefont {Corrado},\ and\ \citenamefont
  {Dean}}]{Word2Vec:13}%
  \BibitemOpen
  \bibfield  {author} {\bibinfo {author} {\bibfnamefont {T.}~\bibnamefont
  {Mikolov}}, \bibinfo {author} {\bibfnamefont {K.}~\bibnamefont {Chen}},
  \bibinfo {author} {\bibfnamefont {G.}~\bibnamefont {Corrado}}, \ and\
  \bibinfo {author} {\bibfnamefont {J.}~\bibnamefont {Dean}},\ }\href@noop {}
  {\enquote {\bibinfo {title} {Efficient estimation of word representations in
  vector space},}\ } (\bibinfo {year} {2013}),\ \Eprint
  {http://arxiv.org/abs/1301.3781} {arXiv:1301.3781} \BibitemShut {NoStop}%
\bibitem [{\citenamefont {Lookman}\ \emph {et~al.}(2019)\citenamefont
  {Lookman}, \citenamefont {Balachandran}, \citenamefont {Xue},\ and\
  \citenamefont {Yuan}}]{Lookman-AL:19}%
  \BibitemOpen
  \bibfield  {author} {\bibinfo {author} {\bibfnamefont {T.}~\bibnamefont
  {Lookman}}, \bibinfo {author} {\bibfnamefont {P.~V.}\ \bibnamefont
  {Balachandran}}, \bibinfo {author} {\bibfnamefont {D.}~\bibnamefont {Xue}}, \
  and\ \bibinfo {author} {\bibfnamefont {R.}~\bibnamefont {Yuan}},\ }\href
  {\doibase 10.1038/s41524-019-0153-8} {\bibfield  {journal} {\bibinfo
  {journal} {npj Comput. Mater.}\ }\textbf {\bibinfo {volume} {5}},\ \bibinfo
  {pages} {21} (\bibinfo {year} {2019})}\BibitemShut {NoStop}%
\bibitem [{\citenamefont {Kohn}\ and\ \citenamefont {Sham}(1965)}]{KoSh65}%
  \BibitemOpen
  \bibfield  {author} {\bibinfo {author} {\bibfnamefont {W.}~\bibnamefont
  {Kohn}}\ and\ \bibinfo {author} {\bibfnamefont {L.~J.}\ \bibnamefont
  {Sham}},\ }\href {\doibase 10.1103/PhysRev.140.A1133} {\bibfield  {journal}
  {\bibinfo  {journal} {Phys. Rev.}\ }\textbf {\bibinfo {volume} {140}},\
  \bibinfo {pages} {A1133} (\bibinfo {year} {1965})}\BibitemShut {NoStop}%
\bibitem [{\citenamefont {Kresse}\ and\ \citenamefont
  {Joubert}(1999)}]{USPP-PAW:99}%
  \BibitemOpen
  \bibfield  {author} {\bibinfo {author} {\bibfnamefont {G.}~\bibnamefont
  {Kresse}}\ and\ \bibinfo {author} {\bibfnamefont {D.}~\bibnamefont
  {Joubert}},\ }\href {\doibase 10.1103/PhysRevB.59.1758} {\bibfield  {journal}
  {\bibinfo  {journal} {Phys. Rev. B}\ }\textbf {\bibinfo {volume} {59}},\
  \bibinfo {pages} {1758} (\bibinfo {year} {1999})}\BibitemShut {NoStop}%
\bibitem [{\citenamefont {Bl\"ochl}(1994)}]{PAW:94}%
  \BibitemOpen
  \bibfield  {author} {\bibinfo {author} {\bibfnamefont {P.~E.}\ \bibnamefont
  {Bl\"ochl}},\ }\href {\doibase 10.1103/PhysRevB.50.17953} {\bibfield
  {journal} {\bibinfo  {journal} {Phys. Rev. B}\ }\textbf {\bibinfo {volume}
  {50}},\ \bibinfo {pages} {17953} (\bibinfo {year} {1994})}\BibitemShut
  {NoStop}%
\bibitem [{\citenamefont {Perdew}\ \emph {et~al.}(1996)\citenamefont {Perdew},
  \citenamefont {Burke},\ and\ \citenamefont {Ernzerhof}}]{PeBu96}%
  \BibitemOpen
  \bibfield  {author} {\bibinfo {author} {\bibfnamefont {J.~P.}\ \bibnamefont
  {Perdew}}, \bibinfo {author} {\bibfnamefont {K.}~\bibnamefont {Burke}}, \
  and\ \bibinfo {author} {\bibfnamefont {M.}~\bibnamefont {Ernzerhof}},\ }\href
  {\doibase 10.1103/PhysRevLett.77.3865} {\bibfield  {journal} {\bibinfo
  {journal} {Phys. Rev. Lett.}\ }\textbf {\bibinfo {volume} {77}},\ \bibinfo
  {pages} {3865} (\bibinfo {year} {1996})}\BibitemShut {NoStop}%
\bibitem [{\citenamefont {Ong}\ \emph {et~al.}(2013)\citenamefont {Ong},
  \citenamefont {Richards}, \citenamefont {Jain}, \citenamefont {Hautier},
  \citenamefont {Kocher}, \citenamefont {Cholia}, \citenamefont {Gunter},
  \citenamefont {Chevrier}, \citenamefont {Persson},\ and\ \citenamefont
  {Ceder}}]{PYMATGEN:13}%
  \BibitemOpen
  \bibfield  {author} {\bibinfo {author} {\bibfnamefont {S.~P.}\ \bibnamefont
  {Ong}}, \bibinfo {author} {\bibfnamefont {W.~D.}\ \bibnamefont {Richards}},
  \bibinfo {author} {\bibfnamefont {A.}~\bibnamefont {Jain}}, \bibinfo {author}
  {\bibfnamefont {G.}~\bibnamefont {Hautier}}, \bibinfo {author} {\bibfnamefont
  {M.}~\bibnamefont {Kocher}}, \bibinfo {author} {\bibfnamefont
  {S.}~\bibnamefont {Cholia}}, \bibinfo {author} {\bibfnamefont
  {D.}~\bibnamefont {Gunter}}, \bibinfo {author} {\bibfnamefont {V.~L.}\
  \bibnamefont {Chevrier}}, \bibinfo {author} {\bibfnamefont {K.~A.}\
  \bibnamefont {Persson}}, \ and\ \bibinfo {author} {\bibfnamefont
  {G.}~\bibnamefont {Ceder}},\ }\href {\doibase
  https://doi.org/10.1016/j.commatsci.2012.10.028} {\bibfield  {journal}
  {\bibinfo  {journal} {Comp. Mat. Sci.}\ }\textbf {\bibinfo {volume} {68}},\
  \bibinfo {pages} {314} (\bibinfo {year} {2013})}\BibitemShut {NoStop}%
\bibitem [{\citenamefont {Liechtenstein}\ \emph {et~al.}(1995)\citenamefont
  {Liechtenstein}, \citenamefont {Anisimov},\ and\ \citenamefont
  {Zaanen}}]{LiechtensteinAnisimov:95}%
  \BibitemOpen
  \bibfield  {author} {\bibinfo {author} {\bibfnamefont {A.~I.}\ \bibnamefont
  {Liechtenstein}}, \bibinfo {author} {\bibfnamefont {V.~I.}\ \bibnamefont
  {Anisimov}}, \ and\ \bibinfo {author} {\bibfnamefont {J.}~\bibnamefont
  {Zaanen}},\ }\href {\doibase 10.1103/PhysRevB.52.R5467} {\bibfield  {journal}
  {\bibinfo  {journal} {Phys. Rev. B}\ }\textbf {\bibinfo {volume} {52}},\
  \bibinfo {pages} {R5467} (\bibinfo {year} {1995})}\BibitemShut {NoStop}%
\bibitem [{\citenamefont {Liu}\ \emph {et~al.}(2013)\citenamefont {Liu},
  \citenamefont {Kargarian}, \citenamefont {Kareev}, \citenamefont {Gray},
  \citenamefont {Ryan}, \citenamefont {Cruz}, \citenamefont {Tahir},
  \citenamefont {Chuang}, \citenamefont {Guo}, \citenamefont {Rondinelli},
  \citenamefont {Freeland}, \citenamefont {Fiete},\ and\ \citenamefont
  {Chakhalian}}]{Liu-NNO:13}%
  \BibitemOpen
  \bibfield  {author} {\bibinfo {author} {\bibfnamefont {J.}~\bibnamefont
  {Liu}}, \bibinfo {author} {\bibfnamefont {M.}~\bibnamefont {Kargarian}},
  \bibinfo {author} {\bibfnamefont {M.}~\bibnamefont {Kareev}}, \bibinfo
  {author} {\bibfnamefont {B.}~\bibnamefont {Gray}}, \bibinfo {author}
  {\bibfnamefont {P.~J.}\ \bibnamefont {Ryan}}, \bibinfo {author}
  {\bibfnamefont {A.}~\bibnamefont {Cruz}}, \bibinfo {author} {\bibfnamefont
  {N.}~\bibnamefont {Tahir}}, \bibinfo {author} {\bibfnamefont {Y.-D.}\
  \bibnamefont {Chuang}}, \bibinfo {author} {\bibfnamefont {J.}~\bibnamefont
  {Guo}}, \bibinfo {author} {\bibfnamefont {J.~M.}\ \bibnamefont {Rondinelli}},
  \bibinfo {author} {\bibfnamefont {J.~W.}\ \bibnamefont {Freeland}}, \bibinfo
  {author} {\bibfnamefont {G.~A.}\ \bibnamefont {Fiete}}, \ and\ \bibinfo
  {author} {\bibfnamefont {J.}~\bibnamefont {Chakhalian}},\ }\href {\doibase
  10.1038/ncomms3714} {\bibfield  {journal} {\bibinfo  {journal} {Nat.
  Commun.}\ }\textbf {\bibinfo {volume} {4}},\ \bibinfo {pages} {2714}
  (\bibinfo {year} {2013})}\BibitemShut {NoStop}%
\bibitem [{\citenamefont {Abadi}\ \emph {et~al.}(2015)\citenamefont {Abadi},
  \citenamefont {Agarwal}, \citenamefont {Barham}, \citenamefont {Brevdo},
  \citenamefont {Chen}, \citenamefont {Citro}, \citenamefont {Corrado},
  \citenamefont {Davis}, \citenamefont {Dean}, \citenamefont {Devin},
  \citenamefont {Ghemawat}, \citenamefont {Goodfellow}, \citenamefont {Harp},
  \citenamefont {Irving}, \citenamefont {Isard}, \citenamefont {Jia},
  \citenamefont {Jozefowicz}, \citenamefont {Kaiser}, \citenamefont {Kudlur},
  \citenamefont {Levenberg}, \citenamefont {Man\'{e}}, \citenamefont {Monga},
  \citenamefont {Moore}, \citenamefont {Murray}, \citenamefont {Olah},
  \citenamefont {Schuster}, \citenamefont {Shlens}, \citenamefont {Steiner},
  \citenamefont {Sutskever}, \citenamefont {Talwar}, \citenamefont {Tucker},
  \citenamefont {Vanhoucke}, \citenamefont {Vasudevan}, \citenamefont
  {Vi\'{e}gas}, \citenamefont {Vinyals}, \citenamefont {Warden}, \citenamefont
  {Wattenberg}, \citenamefont {Wicke}, \citenamefont {Yu},\ and\ \citenamefont
  {Zheng}}]{Tensorflow:15}%
  \BibitemOpen
  \bibfield  {author} {\bibinfo {author} {\bibfnamefont {M.}~\bibnamefont
  {Abadi}}, \bibinfo {author} {\bibfnamefont {A.}~\bibnamefont {Agarwal}},
  \bibinfo {author} {\bibfnamefont {P.}~\bibnamefont {Barham}}, \bibinfo
  {author} {\bibfnamefont {E.}~\bibnamefont {Brevdo}}, \bibinfo {author}
  {\bibfnamefont {Z.}~\bibnamefont {Chen}}, \bibinfo {author} {\bibfnamefont
  {C.}~\bibnamefont {Citro}}, \bibinfo {author} {\bibfnamefont {G.~S.}\
  \bibnamefont {Corrado}}, \bibinfo {author} {\bibfnamefont {A.}~\bibnamefont
  {Davis}}, \bibinfo {author} {\bibfnamefont {J.}~\bibnamefont {Dean}},
  \bibinfo {author} {\bibfnamefont {M.}~\bibnamefont {Devin}}, \bibinfo
  {author} {\bibfnamefont {S.}~\bibnamefont {Ghemawat}}, \bibinfo {author}
  {\bibfnamefont {I.}~\bibnamefont {Goodfellow}}, \bibinfo {author}
  {\bibfnamefont {A.}~\bibnamefont {Harp}}, \bibinfo {author} {\bibfnamefont
  {G.}~\bibnamefont {Irving}}, \bibinfo {author} {\bibfnamefont
  {M.}~\bibnamefont {Isard}}, \bibinfo {author} {\bibfnamefont
  {Y.}~\bibnamefont {Jia}}, \bibinfo {author} {\bibfnamefont {R.}~\bibnamefont
  {Jozefowicz}}, \bibinfo {author} {\bibfnamefont {L.}~\bibnamefont {Kaiser}},
  \bibinfo {author} {\bibfnamefont {M.}~\bibnamefont {Kudlur}}, \bibinfo
  {author} {\bibfnamefont {J.}~\bibnamefont {Levenberg}}, \bibinfo {author}
  {\bibfnamefont {D.}~\bibnamefont {Man\'{e}}}, \bibinfo {author}
  {\bibfnamefont {R.}~\bibnamefont {Monga}}, \bibinfo {author} {\bibfnamefont
  {S.}~\bibnamefont {Moore}}, \bibinfo {author} {\bibfnamefont
  {D.}~\bibnamefont {Murray}}, \bibinfo {author} {\bibfnamefont
  {C.}~\bibnamefont {Olah}}, \bibinfo {author} {\bibfnamefont {M.}~\bibnamefont
  {Schuster}}, \bibinfo {author} {\bibfnamefont {J.}~\bibnamefont {Shlens}},
  \bibinfo {author} {\bibfnamefont {B.}~\bibnamefont {Steiner}}, \bibinfo
  {author} {\bibfnamefont {I.}~\bibnamefont {Sutskever}}, \bibinfo {author}
  {\bibfnamefont {K.}~\bibnamefont {Talwar}}, \bibinfo {author} {\bibfnamefont
  {P.}~\bibnamefont {Tucker}}, \bibinfo {author} {\bibfnamefont
  {V.}~\bibnamefont {Vanhoucke}}, \bibinfo {author} {\bibfnamefont
  {V.}~\bibnamefont {Vasudevan}}, \bibinfo {author} {\bibfnamefont
  {F.}~\bibnamefont {Vi\'{e}gas}}, \bibinfo {author} {\bibfnamefont
  {O.}~\bibnamefont {Vinyals}}, \bibinfo {author} {\bibfnamefont
  {P.}~\bibnamefont {Warden}}, \bibinfo {author} {\bibfnamefont
  {M.}~\bibnamefont {Wattenberg}}, \bibinfo {author} {\bibfnamefont
  {M.}~\bibnamefont {Wicke}}, \bibinfo {author} {\bibfnamefont
  {Y.}~\bibnamefont {Yu}}, \ and\ \bibinfo {author} {\bibfnamefont
  {X.}~\bibnamefont {Zheng}},\ }\href {https://www.tensorflow.org/} {\enquote
  {\bibinfo {title} {{TensorFlow}: Large-scale machine learning on
  heterogeneous systems},}\ } (\bibinfo {year} {2015}),\ \bibinfo {note}
  {software available from tensorflow.org}\BibitemShut {NoStop}%
\bibitem [{\citenamefont {Chollet}(2015)}]{CholletKeras:15}%
  \BibitemOpen
  \bibfield  {author} {\bibinfo {author} {\bibfnamefont {F.}~\bibnamefont
  {Chollet}},\ }\href@noop {} {\enquote {\bibinfo {title} {Keras},}\ }\bibinfo
  {howpublished} {\url{https://keras.io}} (\bibinfo {year} {2015})\BibitemShut
  {NoStop}%
\bibitem [{Note1()}]{Note1}%
  \BibitemOpen
  \bibinfo {note} {The well-known overbinding of gas-phase O$_2$ molecules in
  DFT necessitates a correction of $E(\protect \text {O$_2$})$, which we
  performed such as to reproduce the experimental O$_2$ binding energy of
  $5.16$~eV~\cite {GeislerPentcheva-LNOLAO-Resonances:19,
  GeislerPentcheva-LCO:20, LNO-OxVac-Beigi:15}.}\BibitemShut {Stop}%
\bibitem [{\citenamefont {Nestola}\ \emph {et~al.}(2018)\citenamefont
  {Nestola}, \citenamefont {Korolev}, \citenamefont {Kopylova}, \citenamefont
  {Rotiroti}, \citenamefont {Pearson}, \citenamefont {Pamato}, \citenamefont
  {Alvaro}, \citenamefont {Peruzzo}, \citenamefont {Gurney}, \citenamefont
  {Moore},\ and\ \citenamefont {Davidson}}]{Nestola-CaSiO3Perov:18}%
  \BibitemOpen
  \bibfield  {author} {\bibinfo {author} {\bibfnamefont {F.}~\bibnamefont
  {Nestola}}, \bibinfo {author} {\bibfnamefont {N.}~\bibnamefont {Korolev}},
  \bibinfo {author} {\bibfnamefont {M.}~\bibnamefont {Kopylova}}, \bibinfo
  {author} {\bibfnamefont {N.}~\bibnamefont {Rotiroti}}, \bibinfo {author}
  {\bibfnamefont {D.~G.}\ \bibnamefont {Pearson}}, \bibinfo {author}
  {\bibfnamefont {M.~G.}\ \bibnamefont {Pamato}}, \bibinfo {author}
  {\bibfnamefont {M.}~\bibnamefont {Alvaro}}, \bibinfo {author} {\bibfnamefont
  {L.}~\bibnamefont {Peruzzo}}, \bibinfo {author} {\bibfnamefont {J.~J.}\
  \bibnamefont {Gurney}}, \bibinfo {author} {\bibfnamefont {A.~E.}\
  \bibnamefont {Moore}}, \ and\ \bibinfo {author} {\bibfnamefont
  {J.}~\bibnamefont {Davidson}},\ }\href {\doibase 10.1038/nature25972}
  {\bibfield  {journal} {\bibinfo  {journal} {Nature}\ }\textbf {\bibinfo
  {volume} {555}},\ \bibinfo {pages} {237} (\bibinfo {year}
  {2018})}\BibitemShut {NoStop}%
\bibitem [{\citenamefont {Kirklin}\ \emph {et~al.}(2015)\citenamefont
  {Kirklin}, \citenamefont {Saal}, \citenamefont {Meredig}, \citenamefont
  {Thompson}, \citenamefont {Doak}, \citenamefont {Aykol}, \citenamefont
  {R{\"u}hl},\ and\ \citenamefont {Wolverton}}]{KirklinWolverton-DFTvsAI:15}%
  \BibitemOpen
  \bibfield  {author} {\bibinfo {author} {\bibfnamefont {S.}~\bibnamefont
  {Kirklin}}, \bibinfo {author} {\bibfnamefont {J.~E.}\ \bibnamefont {Saal}},
  \bibinfo {author} {\bibfnamefont {B.}~\bibnamefont {Meredig}}, \bibinfo
  {author} {\bibfnamefont {A.}~\bibnamefont {Thompson}}, \bibinfo {author}
  {\bibfnamefont {J.~W.}\ \bibnamefont {Doak}}, \bibinfo {author}
  {\bibfnamefont {M.}~\bibnamefont {Aykol}}, \bibinfo {author} {\bibfnamefont
  {S.}~\bibnamefont {R{\"u}hl}}, \ and\ \bibinfo {author} {\bibfnamefont
  {C.}~\bibnamefont {Wolverton}},\ }\href {\doibase
  10.1038/npjcompumats.2015.10} {\bibfield  {journal} {\bibinfo  {journal} {npj
  Comput. Mater.}\ }\textbf {\bibinfo {volume} {1}},\ \bibinfo {pages} {15010}
  (\bibinfo {year} {2015})}\BibitemShut {NoStop}%
\bibitem [{\citenamefont {van~der Maaten}\ and\ \citenamefont
  {Hinton}(2008)}]{tSNE:08}%
  \BibitemOpen
  \bibfield  {author} {\bibinfo {author} {\bibfnamefont {L.}~\bibnamefont
  {van~der Maaten}}\ and\ \bibinfo {author} {\bibfnamefont {G.}~\bibnamefont
  {Hinton}},\ }\href {http://jmlr.org/papers/v9/vandermaaten08a.html}
  {\bibfield  {journal} {\bibinfo  {journal} {J. Mach. Learn. Res.}\ }\textbf
  {\bibinfo {volume} {9}},\ \bibinfo {pages} {2579} (\bibinfo {year}
  {2008})}\BibitemShut {NoStop}%
\bibitem [{\citenamefont {Geisler}\ \emph {et~al.}(2017)\citenamefont
  {Geisler}, \citenamefont {Blanca-Romero},\ and\ \citenamefont
  {Pentcheva}}]{Geisler-LNOSTO:17}%
  \BibitemOpen
  \bibfield  {author} {\bibinfo {author} {\bibfnamefont {B.}~\bibnamefont
  {Geisler}}, \bibinfo {author} {\bibfnamefont {A.}~\bibnamefont
  {Blanca-Romero}}, \ and\ \bibinfo {author} {\bibfnamefont {R.}~\bibnamefont
  {Pentcheva}},\ }\href {\doibase 10.1103/PhysRevB.95.125301} {\bibfield
  {journal} {\bibinfo  {journal} {Phys. Rev. B}\ }\textbf {\bibinfo {volume}
  {95}},\ \bibinfo {pages} {125301} (\bibinfo {year} {2017})}\BibitemShut
  {NoStop}%
\bibitem [{\citenamefont {Geisler}\ and\ \citenamefont
  {Pentcheva}(2018)}]{GeislerPentcheva-LNOLAO:18}%
  \BibitemOpen
  \bibfield  {author} {\bibinfo {author} {\bibfnamefont {B.}~\bibnamefont
  {Geisler}}\ and\ \bibinfo {author} {\bibfnamefont {R.}~\bibnamefont
  {Pentcheva}},\ }\href {\doibase 10.1103/PhysRevMaterials.2.055403} {\bibfield
   {journal} {\bibinfo  {journal} {Phys. Rev. Materials}\ }\textbf {\bibinfo
  {volume} {2}},\ \bibinfo {pages} {055403} (\bibinfo {year}
  {2018})}\BibitemShut {NoStop}%
\bibitem [{\citenamefont {Xing}\ \emph {et~al.}(2017)\citenamefont {Xing},
  \citenamefont {Sun}, \citenamefont {Li}, \citenamefont {Fan}, \citenamefont
  {Zheng},\ and\ \citenamefont {Singh}}]{TE-Fitness-Function-XingSingh:17}%
  \BibitemOpen
  \bibfield  {author} {\bibinfo {author} {\bibfnamefont {G.}~\bibnamefont
  {Xing}}, \bibinfo {author} {\bibfnamefont {J.}~\bibnamefont {Sun}}, \bibinfo
  {author} {\bibfnamefont {Y.}~\bibnamefont {Li}}, \bibinfo {author}
  {\bibfnamefont {X.}~\bibnamefont {Fan}}, \bibinfo {author} {\bibfnamefont
  {W.}~\bibnamefont {Zheng}}, \ and\ \bibinfo {author} {\bibfnamefont {D.~J.}\
  \bibnamefont {Singh}},\ }\href {\doibase 10.1103/PhysRevMaterials.1.065405}
  {\bibfield  {journal} {\bibinfo  {journal} {Phys. Rev. Materials}\ }\textbf
  {\bibinfo {volume} {1}},\ \bibinfo {pages} {065405} (\bibinfo {year}
  {2017})}\BibitemShut {NoStop}%
\bibitem [{\citenamefont {Geisler}\ and\ \citenamefont
  {Pentcheva}(2019)}]{GeislerPentcheva-LNOLAO-Resonances:19}%
  \BibitemOpen
  \bibfield  {author} {\bibinfo {author} {\bibfnamefont {B.}~\bibnamefont
  {Geisler}}\ and\ \bibinfo {author} {\bibfnamefont {R.}~\bibnamefont
  {Pentcheva}},\ }\href {\doibase 10.1103/PhysRevApplied.11.044047} {\bibfield
  {journal} {\bibinfo  {journal} {Phys. Rev. Applied}\ }\textbf {\bibinfo
  {volume} {11}},\ \bibinfo {pages} {044047} (\bibinfo {year}
  {2019})}\BibitemShut {NoStop}%
\bibitem [{\citenamefont {Geisler}\ and\ \citenamefont
  {Pentcheva}(2020{\natexlab{b}})}]{GeislerPentcheva-LCO:20}%
  \BibitemOpen
  \bibfield  {author} {\bibinfo {author} {\bibfnamefont {B.}~\bibnamefont
  {Geisler}}\ and\ \bibinfo {author} {\bibfnamefont {R.}~\bibnamefont
  {Pentcheva}},\ }\href {\doibase 10.1103/PhysRevB.101.165108} {\bibfield
  {journal} {\bibinfo  {journal} {Phys. Rev. B}\ }\textbf {\bibinfo {volume}
  {101}},\ \bibinfo {pages} {165108} (\bibinfo {year}
  {2020}{\natexlab{b}})}\BibitemShut {NoStop}%
\bibitem [{\citenamefont {Malashevich}\ and\ \citenamefont
  {Ismail-Beigi}(2015)}]{LNO-OxVac-Beigi:15}%
  \BibitemOpen
  \bibfield  {author} {\bibinfo {author} {\bibfnamefont {A.}~\bibnamefont
  {Malashevich}}\ and\ \bibinfo {author} {\bibfnamefont {S.}~\bibnamefont
  {Ismail-Beigi}},\ }\href {\doibase 10.1103/PhysRevB.92.144102} {\bibfield
  {journal} {\bibinfo  {journal} {Phys. Rev. B}\ }\textbf {\bibinfo {volume}
  {92}},\ \bibinfo {pages} {144102} (\bibinfo {year} {2015})}\BibitemShut
  {NoStop}%
\end{thebibliography}

%

\end{document}